\documentclass[aip,reprint]{revtex4-1}
\usepackage{multirow}%
\usepackage[compact]{titlesec}
\titlespacing{\section}{0pt}{3ex}{3ex}
\titlespacing{\subsection}{0pt}{2ex}{2ex}
\titlespacing{\subsubsection}{0pt}{1ex}{1ex}
\usepackage{graphicx}
\usepackage{dcolumn}
\usepackage{bm}
\usepackage{float}

\usepackage[font=small,labelfont=bf,
   justification=Justified,
   format=plain,belowskip= 0 pt,aboveskip=0 pt]{caption} 
\usepackage{subcaption}
\usepackage[dvipsnames]{xcolor}
\usepackage{comment}
\usepackage{xfrac}
\usepackage{amsmath, amssymb,bbding,oplotsymbl,MnSymbol}
\usepackage{threeparttable}

\usepackage{hyperref}
\hypersetup{
    colorlinks=true,
    linkcolor=orange,
    citecolor=red,
    urlcolor=blue
}
\usepackage{bm}    
\DeclareSymbolFont{symbolsC}{U}{pxsyc}{m}{n}
\SetSymbolFont{symbolsC}{bold}{U}{pxsyc}{bx}{n}
\DeclareUnicodeCharacter{2212}{-}
\DeclareFontSubstitution{U}{pxsyc}{m}{n}
\DeclareMathSymbol{\medcirc}{\mathbin}{symbolsC}{7}
\DeclareFontSubstitution{OMS}{pxsy}{m}{n}

\definecolor{maroon}{rgb}{0.64, 0.08, 0.18}	

\definecolor{orange1}{rgb}{0.99, 0.50, 0.07}	
\definecolor{blue1}{rgb}{0.01, 0.21, 1.0}	
\definecolor{pink1}{rgb}{0.93, 0.01, 0.99}	
\definecolor{purple1}{rgb}{0.40, 0.02, 0.98}	
\definecolor{myPurple}{rgb}{0.5,0,0.5}

\definecolor{d_cyan}{rgb}{0.0, 0.45,0.74}

\usepackage{lineno}

\newcommand{\RomanNumeralCaps}[1]
\draft 

\begin{document}


\title{Inertial effects on flow dynamics near a moving contact line} 



\author{Charul Gupta}
 \email{charul.gupta229@gmail.com}
\affiliation{Department of Mechanical \& Aerospace Engineering, Indian Institute of Technology Hyderabad, India\\
}%

\author{Rishabh Sharma}
 \email{me22mtech12009@iith.ac.in}
\affiliation{Department of Mechanical \& Aerospace Engineering, Indian Institute of Technology Hyderabad, India\\
}%

\author{Tejasvi Hegde}
 \email{me22mtech12006@iith.ac.in}
\affiliation{Department of Mechanical \& Aerospace Engineering, Indian Institute of Technology Hyderabad, India\\
}%

\author{Venkata Sai Anvesh Sangadi}
 \email{me20m22p100001@iith.ac.in}
\affiliation{Department of Mechanical \& Aerospace Engineering, Indian Institute of Technology Hyderabad, India\\
}%

\author{Lakshmana Dora Chandrala}
\email{lchandrala@mae.iith.ac.in}
\affiliation{Department of Mechanical \& Aerospace Engineering, Indian Institute of Technology Hyderabad, India\\
}%

\author{Harish N Dixit}
\email{hdixit@mae.iith.ac.in}
\affiliation{Department of Mechanical \& Aerospace Engineering, Indian Institute of Technology Hyderabad, India\\
}%
\affiliation{Centre for Interdisciplinary Programs, Indian Institute of Technology Hyderabad, India\\
}%




\date{\today}

\begin{abstract}
This study investigates the role of inertia in moving contact lines using experiments, theoretical analysis, and numerical simulations. Experiments are conducted using a plate immersion configuration over a wide range of Reynolds numbers from $O(10^{-3})$ to $O(10)$. Flow configurations and quantitative measurements are obtained using high-speed imaging and particle image velocimetry. The streamfunction contours reconstructed from the experimental velocity fields are compared with the viscous modulated wedge solution (viscous-MWS) and inertial-MWS theory. Experimental observations show that the streamfunction contours agree well with viscous predictions at low Reynolds numbers; however, systematic deviations emerge as the Reynolds number increases. The inertial-MWS theory, an inertial extension of the Huh and Scriven framework, accounts for these deviations, but only within a narrow range of Reynolds numbers $10^{-1} < Re < 1$. At higher Reynolds numbers, inertial theory fails to accurately capture the deviations in the streamfunction contours observed in the experiments. Moreover, simulations conducted using the volume of fluid method support our findings, exhibiting deviations in streamfunction contours consistent with experimental observations. We demonstrate that inertia does not fundamentally alter the underlying flow configuration but instead induces a systematic deviation in the streamfunction contours. At finite $Re$, the interfacial speed transitions from a nearly constant value in the viscous regime to a monotonic decay along the interface. These findings expose the need for more sophisticated models of moving contact lines. 
\end{abstract}


\maketitle 

\section{Introduction}\label{sec:intro}
Fluids flowing over solid surfaces are common in both natural and industrial settings. This phenomenon occurs in a wide range of applications, including painting and coating processes, immersion lithography \cite{he_eindhoven2020}, and multiphase flow heat transfer. A moving contact line is a common phenomenon underlying all these applications, which makes understanding the governing physics in its vicinity critically important. The moving contact line typically occurs when an interface between two immiscible fluids advances over a solid surface. 
The motion of the contact line over surfaces can span a wide range of Reynolds numbers, from $O(10^{-3})$ for highly viscous fluids to $O(10^2)$ for low viscosity fluids such as water. This broad range of Reynolds numbers, as well as the inherent multiscale and multiphysics nature \cite{snoeijer2013moving} of the phenomenon, have spurred extensive research on moving contact lines. Most real-world scenarios \cite{duez2007making} occur at high Reynolds numbers, making inertia an inherent part of moving contact lines. Despite its importance, inertia near moving contact lines has received little attention. The objective of the present study was to investigate the moving contact line phenomenon, particularly at high Reynolds numbers ($Re>1$), to advance our understanding of inertial effects near the moving contact line.

The moving contact line phenomenon has been investigated mainly in the viscous regime ($Re = \rho U L/\mu \ll 1$) through multiple approaches, such as theoretical models, experimental observations, and computations. Here, $\rho$, $U$, $L$, and $\mu$ denote the density of the fluid, the speed of the solid, the macroscopic length scale of the system, and the dynamic viscosity of the fluid, respectively. A simple theoretical model by Huh $\&$ Scriven \cite{huh1971hydrodynamic} (HS71 hereafter) reported that only two parameters: the viscosity ratio, $\lambda = \mu_A / \mu_B$ and the dynamic contact angle, $\theta_d$ govern the physics near the moving contact line. 
Here, $\mu_A$ and $\mu_B$ denote the dynamic viscosity of the upper and lower fluid phases, respectively. Several studies have attempted to establish a relationship between the contact angle and a contact line speed or a capillary number ($Ca = \mu U/ \sigma$). Here, $\sigma$ denotes the surface tension for an air-liquid interface. The classic experimental study by Hoffmann \cite{hoffman1975study} established that the dynamic contact angle exhibits a universal dependence on the capillary number and the static contact angle ($\theta_s$).  Theoretical models \cite{blake1969kinetics,voinov1976hydrodynamics, cox1986dynamics,shikhmurzaev1997moving}, such as the Cox model, show that the dynamic contact angle varies with the capillary number, the static contact angle, and the ratio of length scales, $L/l_s$. For example, the Cox model \cite{cox1986dynamics} predicts the dynamic contact angle $\theta_d$ as 
$\theta_d^3 - \theta_s^3 \sim Ca \ln(L/l_s)$. Here, $l_s$ denotes the slip length. Comprehensive discussions of contact angle models and moving contact lines can be found in several excellent reviews and monographs \cite{dussan1979spreading,shikhmurzaev2007capillary,snoeijer2013moving}. Most studies on dynamic contact angle have been conducted mainly in the viscous regime \cite{le2005shape,rio2005boundary}, except the study by Puthenveettil \textit{et al.} \cite{puthenveettil2013motion} who conducted high Reynolds number experiments on a drop moving over an inclined plane. They carried out a rigorous comparison between existing contact angle models and their experimentally measured dynamic contact angles. 
 
The contact angle regulates the interface shape, which is inherently curved near the contact line. The interface shape, governed by capillary and gravitational forces under static conditions, becomes more complex under dynamic regimes due to both viscous and inertial effects. The study by Dussan \textit{et al.} \cite{dussan1991} proposed the interface shape model (DRG model), which predicts the dynamic interface shape at multiple scales near the contact line. To do so, they incorporated the static shape into the Cox's contact angle model \cite{cox1986dynamics}, which accounts only for viscous effects. Recent studies \cite{snoeijer2006free, chan2013hydrodynamics,chan2020cox} proposed the Generalized Lubrication interface shape model (GLM model) based on the lubrication theory on a wedge flow, which predicts the full interface shape at a wide range of contact angles. Both models require a tuning parameter to predict the interface shape, which can only be determined by matching the predicted shapes with those obtained from experiments. Consequently, this restricts the independent applicability of the interface shape models. Experimental studies \cite{gupta2023,gupta2024experimental,gupta2025investigation} in our group have provided interface shapes over a wide range of contact angles and rigorously compared them with predictions from the DRG and GLM models. Inertial effects on interface shapes have also been explored experimentally \cite{stoev1999effects} and theoretically by Cox \cite{cox1998inertial}. Inertial effects on flow configurations near the contact line remain largely unexplored.

The HS71 theory reported physically consistent flow patterns in the vicinity of the contact line by employing a no-slip boundary condition at the solid. However, the theory suffers from a singularity at the contact line due to the divergence of shear stress. This behavior, known as the Huh \& Scriven singularity, arises from enforcing the no-slip boundary condition at the contact line. To address this, multiple studies \cite{cox1986dynamics,shikhmurzaev1993moving,shikhmurzaev1997moving} developed and employed different models, such as slip models or precursor film models at the contact line. For example, the three region model by Cox \cite{cox1986dynamics} employed a constant slip at the solid in an inner region to obtain a singularity free solution in the intermediate and outer regions. Recent theoretical studies \cite{kirkinis2013hydrodynamic, kirkinis2014moffatt, febres2017existence} employed a variable slip model at the solid over a finite region and predicted multiple solutions by solving an eigenvalue problem, denoted by different ‘n’ values. Each ‘n’ represents a distinct flow configuration, which makes it challenging to identify the flow structure that would actually occur in reality. In contrast, HS71 theory uniquely represents the flow configurations and is found to be identical to the Cox solution in the intermediate region \cite{cox1986dynamics, shikhmurzaev1997moving, gupta2025investigation}. Earlier studies in our group have rigorously tested the HS71 solution for the flow configuration through experiments \cite{gupta2023, gupta2024experimental, gupta2025investigation} at a wide range of viscosity ratios and dynamic contact angles. These experimental studies have been conducted primarily in the viscous regime. 
To the best of our knowledge, inertial effects near the contact line have not been reported in the literature, with the exception of theoretical studies \cite{hancock1981effects,varma2021inertial}, which lack experimental validation. 
For example, Varma et al. \cite{varma2021inertial} predicted inertial effects on flow configuration by incorporating leading-order inertial terms into the HS71 hydrodynamic model. The goal of the present study is to demonstrate inertial effects near the contact line and to test the theoretical models through direct comparison with experiments to assess their limitations.

The present study reports experiments over a wide range of Reynolds numbers, $10^{-3}<Re<10$, to clearly identify inertial effects near the moving contact line. To date, inertial effects near the moving contact line have only been predicted by theoretical models \cite{hancock1981effects,varma2021inertial}, with no experimental evidence reported. These models incorporate inertia into the HS71 framework as a leading-order perturbative correction, typically using the local Reynolds number as the expansion parameter. Previous experimental studies investigated inertial effects, particularly on the dynamic contact angle and interface shape. The study by our group \cite{gupta2025investigation} differed by focusing primarily on flow dynamics at obtuse contact angles in the viscous regime, while briefly addressing inertial effects.
The present study bridges this gap by providing direct evidence of inertial effects on flow configurations near the contact line through a comparison of experimental data with predictions from viscous and inertia-corrected models. Moreover, simulations have also been carried out to support the experimental evidence on inertial effects. These simulations follow the framework proposed by Fullana \textit{et al.}\cite{fullana2024consistent}, while employing a Navier slip model with a spatially varying slip coefficient. In a recent study \cite{gupta2025investigation} conducted in our group, we reported numerical simulations and directly compared the results with experiments and theory. In the present study, a similar established approach is adopted to probe inertial effects within the flow domain. The findings of the present study, including interface shapes and interfacial speeds, can be directly used to support the development of more accurate predictive models.

The rest of the paper is organized as follows. In \S\ref{sec:experimental_methodology}, we describe the experimental setup, cleaning protocols, and data analysis. In \S\ref{sec:theory_background}, we revisit the theoretical model of Varma \textit{et al.} and develop a framework to facilitate comparison among theory, experiments, and simulations. 
The results on flow fields and interfacial speeds are presented in \S\ref{sec:results}, and the paper concludes with a summary of the main findings in \S\ref{sec:conclusion}.
\section{Experimental Methodology}\label{sec:experimental_methodology}
\begin{figure}
\centering
\includegraphics[trim = 0mm -20mm 0mm 0mm, clip, angle=0,width=0.45\textwidth]{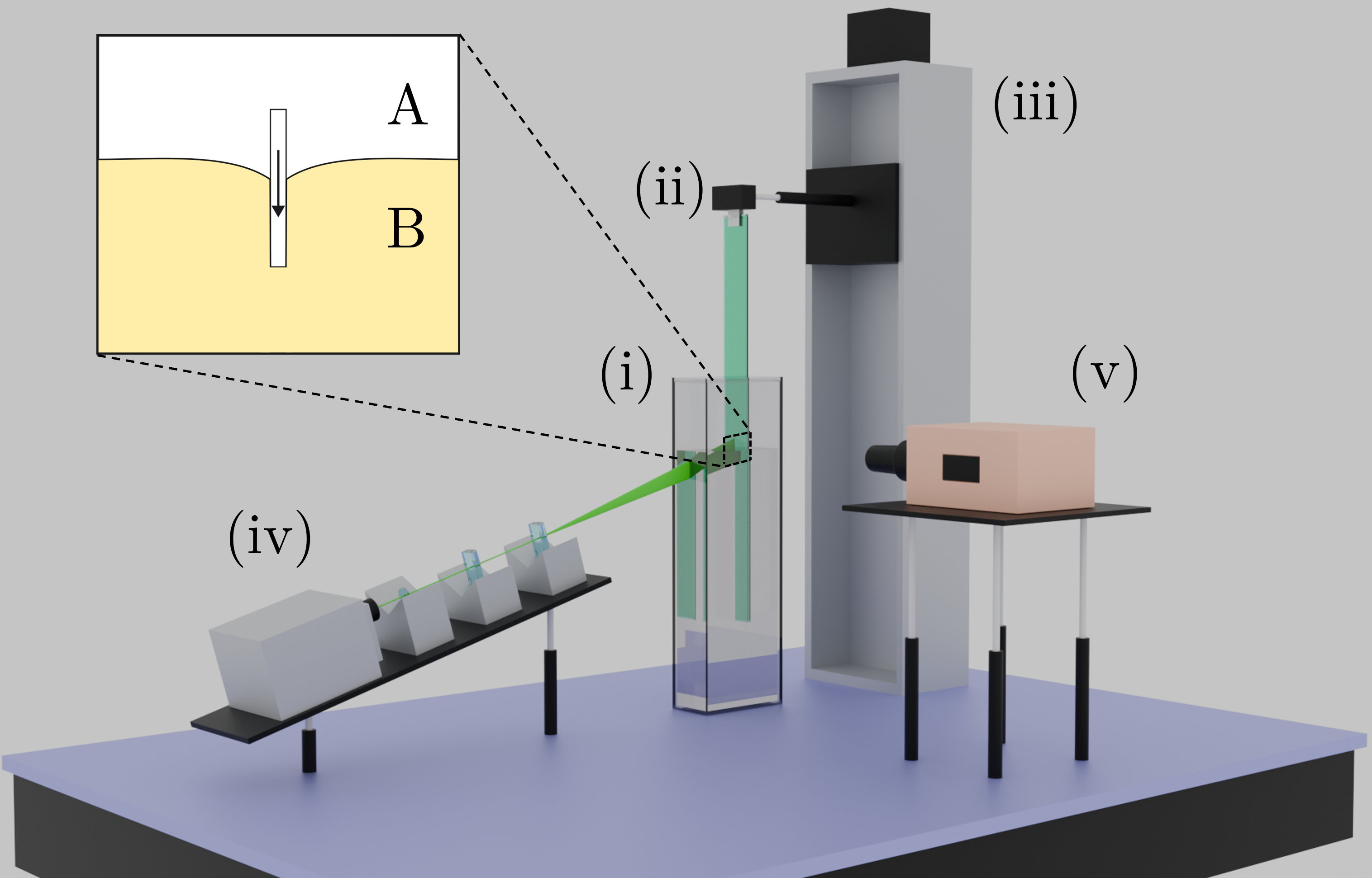}
\caption{ An illustration of the experimental setup. (i) A transparent acrylic tank filled with liquid (ii) A glass plate traversing into the liquid bath (iii) A linear traversing system (iv) A combination of laser and optical lens for forming a thin laser sheet illuminating the region of interest (v) High-speed camera with a macro lens. The inset provides a magnified view of the field of interest, showing the fluid phase B forming an obtuse contact angle with the solid surface.}
\label{fig:experimental_setup}
\end{figure}

A schematic of the experimental setup is shown in Figure \ref{fig:experimental_setup}. It illustrates a custom-built apparatus designed specifically for high Reynolds number experiments.  
The setup comprises the following five key components: (i) a transparent acrylic tank, (ii) a glass plate, (iii) a motorized linear traverse, (iv) an illumination system, and (v) a data acquisition system. The tank, with dimensions $100\text{mm} \times 30\text{mm} \times 100\text{mm}$, was filled with the desired liquid to a height of 80 mm. In this study, air-liquid systems (see Table~\ref{tab:properties_fluids}), such as air-sucrose water mixtures, were used; these systems form a contact angle slightly below $90^{\circ}$ on the surface, which presents challenges for optical visualization. 
Therefore, we coated the tank with a hydrophobic coating to enable clear visualization near the contact line, ensuring a static contact angle greater than $90^{\circ}$.
A glass plate, with dimensions $250\text{mm} \times 25\text{mm} \times 3.5\text{mm}$, was immersed in the tank, which was similarly coated with the hydrophobic coating to maintain an obtuse contact angle. This prevents repeated shifts of the contact angle from acute to obtuse, which could otherwise induce transient motion during the experiments.
The glass plate was attached to a motorized traverse to immerse it into the liquid bath at a controlled speed ranging from 100~$\mu \text{m/s}$ to 20~cm/s (see Table~\ref{tab:operating_parameters}). We selected this range of plate speeds to keep the capillary number $Ca$ below the critical value $Ca^{\text{crit}}$~\cite{vandre2014characteristics}, ensuring that no air entrainment occurs during the experiments. As the plate moves into the liquid, the liquid level rises depending on the dimensions of both the tank and the plate. Changes in the liquid level hinder proper acquisition of particle images and necessitate complex image processing to account for these level shifts. To address this, we used a programmable syringe pump to extract fluid from the bottom of the tank at a rate that corresponded to the increase of the liquid level. This ensured that the interface remained fixed during imaging. We followed cleaning protocols similar to those used in earlier studies~\cite{gupta2023,gupta2024experimental,gupta2025investigation} conducted by our group. Therefore, we highlighted only the critical aspects of the experimental design. 

We mixed the liquid with seeding particles such as 1$\mu \text{m}$ diameter fluorescent polystyrene particles for water-based systems and 5$\mu \text{m}$ diameter particles for silicone oil.
The maximum Stokes number is less than $O (10^{-6})$, indicating that the particles faithfully follow the flow in all of our experiments. We used an illumination system to illuminate the seeding particles in the region near the contact line. The illumination system comprises a 2W 532 nm continuous diode laser, along with a bi-convex spherical lens and a plano-convex cylindrical lens, which together form a thin laser sheet ($\simeq 0.5~ \text{mm}$ thick). The particle images were captured using a high-speed camera (Model: 1MP Photron Nova S9) attached with a macro lens at frame rates of 60-1000 fps depending on the speed of the plate. In all experiments, the field of view was kept 4.2 mm × 4.2 mm, with a spatial resolution of 4$\mu$m/pixel. 

\begin{table*}[t]
  \centering
  \caption{Properties of fluid-fluid system}
  \renewcommand{\arraystretch}{1.1}
  \setlength{\tabcolsep}{6pt}
  \begin{tabular}{lccc}
    \hline
    & \textbf{Density} & \textbf{Viscosity} & \textbf{Surface tension} \\
    & $\rho$           & $\mu$              & $\sigma$           \\
    & kg/m$^3$       & ($10^{-3}$ Pa$\cdot$s) & (mN/m)                \\
    \hline
    Air & 1.207 & 0.0189 & - \\ 
    40\% Sucrose-water mixture & 1190 & 5 & 75.5 \\
    48\% Sugar-water mixture & 1241.8 & 12.3 & 75.5 \\
    50\% Sucrose-water mixture & 1272 & 13.4 & 77.4 \\
    500 cSt Silicone oil & 965 & 516.67 & 18.5 \\
    \hline
\label{tab:properties_fluids}
\end{tabular}
\end{table*}
\begin{table*}[t]
\centering
\begin{threeparttable}
\caption{Operating parameters}
\label{tab:operating_parameters}
\begin{tabular}{ c| @{\hspace{5mm}}c  @{\hspace{5mm}}c  @{\hspace{5mm}}c  @{\hspace{5mm}}c @{\hspace{5mm}}c} 
\hline
Fluid-fluid system &  
\begin{tabular}{@{}c@{}} $U$ (mm/s) \end{tabular} &
\begin{tabular}{@{}c@{}} $Re_{l_c}$ \end{tabular} &
\begin{tabular}{@{}c@{}} $Ca$ \end{tabular} &
\begin{tabular}{@{}c@{}} $\theta_d$ (deg) \end{tabular}  &
\begin{tabular}{@{}c@{}} $Ca^{\text{crit}}$ 
\end{tabular} \\
\hline
\multirow{2}{5cm}{Air - 50\% Sucrose-water} 
& 0.15 & $3.55 \times 10^{-2}$ & $2.6 \times 10^{-5}$ & $\approx 101.2$ & \multirow{2}{1cm}{0.24}\\ 
& 0.2 & $4.73 \times 10^{-2}$ & $3.46 \times 10^{-5}$ & $\approx 102.2$ \\ 
\hline
\multirow{1}{5cm}{Air - 48\% Sugar-water} 
& 0.5 & $1.25 \times 10^{-1}$ & $8.16 \times 10^{-5}$ & $\approx 95.5$ & \multirow{1}{1cm}{0.24}\\ 
\hline
\multirow{4}{5cm}{Air - 40\% Sucrose-water} 
& 5 & $3.02$ & $3.3 \times 10^{-4}$ & $\approx 109.5$ & \multirow{4}{1cm}{0.19} \\ 
& 10 & $6.04$ & $6.6 \times 10^{-4}$ & $\approx 110.1$ \\ 
& 15 & $9.05$ & $9.9 \times 10^{-4}$ & $\approx 112.7$ \\ 
& 20 & $12.1$ & $1.33 \times 10^{-3}$ & $\approx 116.1$ \\
\hline
\multirow{1}{5cm}{Air - 500 cSt Silicone oil} 
& 0.5 & $1.3 \times 10^{-3}$ & $1.4 \times 10^{-2}$ & $\approx 60$ & \multirow{1}{1cm}{0.62}\\ 
\hline
\end{tabular}
\begin{tablenotes}
\scriptsize 
\item[] $U$: Plate speed; $Re_{l_c}$: Reynolds number based on capillary length $l_c$; where $l_c = \sqrt(\sigma/\Delta\rho g)$; $Ca$: Capillary number; $\theta_d$: Dynamic contact angle; 
\item[*] $Ca^{\text{crit}}$: Critical capillary number, calculated using the expression $Ca^{\text{crit}} = 0.123(\mu \text{(in cp units)})^{0.26}$, as reported in \cite{vandre2014characteristics}.
\end{tablenotes}
\end{threeparttable}
\end{table*}

All experiments reached a steady state after the initial transient flow. To distinguish steady-state images from transient ones, we constructed a streakline image by overlaying 300 consecutive images.  
If the streaklines in the image did not intersect, the particle images were considered to correspond to steady flow. The topmost streakline in the image, a locus of the particles at the interface, which extended all the way to the contact line, denoted the interface. The extracted locus was fitted with a two-term exponential function, $y(x)= c_1e^{c_2x} + c_3e^{c_4x}$\cite{gupta2024experimental}, with an $R^2$ of 0.99. We further used this function as input to theoretical models to determine the streamfunction contours, as discussed in detail in \S\ref{sec:wedge_theory}. The experimental interface shapes are presented in terms of the fitting parameters $c_1$, $c_2$, $c_3$ and $c_4$ of the exponential function $y(x)$ in the captions of figures associated with the streamfunction comparison in \S\ref{sec:flowfields}. In experiments, $Re$ was varied from $O(10^{-3})$ to $O(10)$, based on the capillary length $l_c$.

Particle images corresponding to the steady state were preprocessed to enhance contrast and reduce noise prior to PIV analysis. A multigrid window-deforming PIV algorithm was employed and an ensemble-based correlation was used to improve the signal-to-noise ratio in a small interrogation window of 16 pixels $\times$ 16 pixels. The velocity fields obtained from PIV analysis were subsequently used to compute streamfunction contours and interfacial velocities using MATLAB.
\section{Theoretical background}\label{sec:theory_background}
\begin{figure*}
\centering
    \begin{subfigure}[b]{0.25\textwidth}
        \centering
        \includegraphics[width=\textwidth]{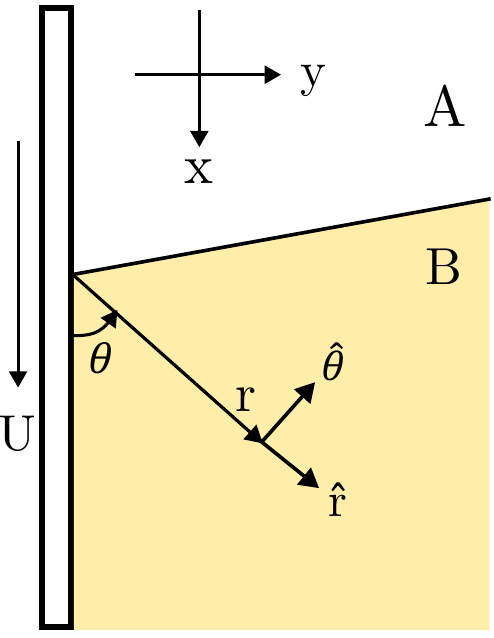}
        \caption{}
        \label{fig:flat_interface}
    \end{subfigure}
    \hspace{30mm}
    \begin{subfigure}[b]{0.25\textwidth}
        \centering
        \includegraphics[width=\textwidth]{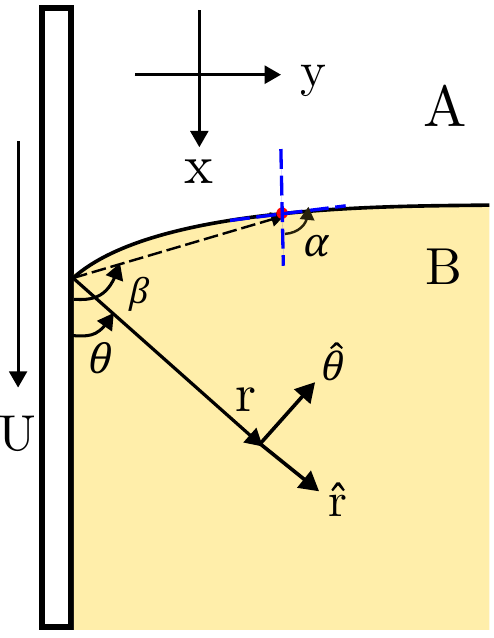}
        \caption{}
        \label{fig:curve_interface}
    \end{subfigure}
\caption{(a) Schematic of geometry showing plate immersion resulting in an advancing contact line. Cylindrical coordinate system ($r$,$\theta$) used for flow near a moving contact line, a flat interface with a constant angle $\phi$, caused by a plate moving at constant speed $U$. (b) Coordinate system for flow near a moving contact line with a curved interface $\beta(r)$.}
\end{figure*}

Several theoretical models have been developed to predict the underlying physics near the moving contact line. A simple theoretical model by HS71 \cite{huh1971hydrodynamic} predicts emerging flow configurations near the moving contact line in the viscous regime. The model is developed assuming a flat interface approximation (see Figure \ref{fig:flat_interface}) and the no-slip boundary condition on the solid. Imposing the no-slip condition resulted in unbounded shear stress at the contact line, giving rise to the long-standing issue known as the Huh \& Scriven singularity. Subsequent theoretical models \cite{cox1986dynamics, shikhmurzaev1993moving,shikhmurzaev1997moving,kirkinis2013hydrodynamic,febres2017existence} were developed primarily to circumvent the singularity at the contact line. For example, the theoretical study by Cox \cite{cox1986dynamics} addressed the singularity by dividing the region near the contact line into three separate regions: (i) an inner region where a constant slip is assumed, (ii) an outer region that depends on the geometry, and (iii) an intermediate region that connects both the inner and outer regions. Using a matched asymptotic expansion method, Cox \cite{cox1986dynamics} reported solutions in the intermediate region. The tangential speed in the inner limit ($u_t|_{r\rightarrow 0}$) was one of the key solutions, representing the interfacial speed for several fluid-fluid systems. A similar solution can be reproduced using the HS71 theoretical framework, indicating that HS71 solution corresponds to the solution in the intermediate region near the moving contact line. This is true even for other theoretical models, such as the model by Shikhmurzaev \textit{et al.} \cite{shikhmurzaev1993moving}. These key insights indicate that testing the HS71 solution is effectively equivalent to testing Cox, Shikhmurzaev, and other related sophisticated models. 

According to the HS71 model, the biharmonic equation in terms of streamfunction ($\psi_v$) governs the physics near the moving contact lines:
\begin{equation}
\label{eqn 3.1}
    \nabla^4 \psi_{v} = 0
\end{equation}
where, $\psi_v$ is expressed in cylindrical polar coordinates as
\begin{equation}
\label{eqn 3.2}
    \psi_{v}(r,\theta) = r f_1(\theta;U)
\end{equation}
Here, $f_1$ is a function of the wedge angle, $\phi$ and the plate speed, $U$. A detailed description of the model is given in the Appendix \ref{apx:viscous-theory}. Solving for the streamfunction subject to the boundary conditions at the solid surface and the interface yields the following expression:
\begin{equation}
\label{eq:streamfunction_vis_p1}
   \psi_v(r,\theta) = r U\left(\frac{\phi\sin\theta - \theta\sin\phi \cos(\theta -\phi) }{\phi - \sin\phi \cos\phi}\right)
\end{equation}
In the above expression, $\psi_v$ denotes the streamfunction in the viscous regime ($r\ll \nu/U$) for a wedge. The expression is obtained by treating fluid phase A as passive, given that it corresponds to air ($\lambda\ll1$). 

\subsection{Inertial theory} \label{sec:inertial-MWS}
Theoretical studies~\cite{hancock1981effects,varma2021inertial} extended the HS71 model to predict inertial effects near the contact line. Instead of solving the full Navier–Stokes equations, which is quite challenging, they adopted a perturbation expansion approach. 
For example, the study by Varma \textit{et al.} \cite{varma2021inertial} employed the perturbation expansion considering the local Reynolds number ($\varrho = r/(\nu/U)<1$, denoted $\varrho$ following
the notation of Varma \textit{et al.}, not to be confused with
fluid density) as a small parameter. Here, $r$ is the distance from the contact line. $\nu$ represents the kinematic viscosity of the liquid. In their formulation, they considered only the leading-order term to incorporate inertial corrections into the HS71 model \cite{huh1971hydrodynamic}. We rederive the complete mathematical formulation of Varma\textit{ et al.}~\cite{varma2021inertial} by transforming the coordinate system to enable a direct comparison with our experiments. Our coefficients differ from their coefficients because they simplified the formulation by measuring angles with respect to the interface, whereas we reference them from the solid surface. 

The two-dimensional steady Navier–Stokes equations can be expressed in terms of the streamfunction ($\tilde{\psi}$) in cylindrical polar coordinates,
\begin{equation}
\label{gov_eqn_I}
   \nabla^4 \tilde{\psi}=\frac{1}{r}\left(\frac{\partial \tilde{\psi}}{\partial \theta} \frac{\partial\left(\nabla^2 \tilde{\psi}\right)}{\partial r}-\frac{\partial \tilde{\psi}}{\partial r} \frac{\partial\left(\nabla^2 \tilde{\psi}\right)}{\partial \theta}\right)
\end{equation}

The streamfunction $\tilde{\psi}$ captures both viscous and inertial effects. The solution of equn.~\ref{gov_eqn_I} can be assumed in the form of a perturbation expansion in terms of the streamfunction as follows:
\begin{equation} \label{series_soln}
    \tilde{\psi}(r,\theta) = \sum_{n=1}^\infty \psi_{n}(r,\theta), \quad \text{where    } \quad\psi_{n}(r,\theta) = r^n f_n({\theta;U})
\end{equation}
They considered only leading-order inertial correction to the viscous term, i.e., weak inertial effects, therefore $n=2$ in equn.~\ref{series_soln} yields
\begin{equation}
    \tilde{\psi}(r,\theta) = r f_1({\theta;U}) +\frac{r^2}{\nu} f_2({\theta};U)  
    \label{exp_sol_I}
\end{equation}
Here, $\tilde{\psi}$ is a dimensional streamfunction that incorporates a linear combination of the streamfunctions associated with the viscous term and the leading-order inertial term in the flow. For $r \ll \nu/U$, equns.~\ref{gov_eqn_I} and \ref{exp_sol_I} will reduce to the biharmonic equation, i.e. equn.~\ref{eqn 3.1} as the leading-order term will become insignificant compared to the viscous term. For $rU/\nu \leq 1$, the leading-order term is comparable to the viscous term, which is the case explored in the present study. The equn.~\ref{exp_sol_I} can also be expressed as
\begin{equation} \label{eqn:inertial-equation}
    \tilde{\psi}(r,\theta) = \psi_{v}(r,\theta) +\psi_{c}(r,\theta)  
\end{equation}
Where
\begin{equation}
    \psi_{c}(r,\theta) = \frac{r^2}{\nu} f_2({\theta;U})
    \label{psi_cor}
\end{equation}
By substituting the streamfunction from equn.~\ref{exp_sol_I} into the governing equn.~\ref{gov_eqn_I}, we obtain,
\begin{equation}
\label{gensolvf}
    f''''_2 + 4 f''_{2} = -2f_{1}f'_{1} -f'_{1}f''_{1} - f_{1}f'''_{1}
\end{equation}
Equn.~\ref{gensolvf} is a nonhomogeneous linear ordinary differential equation in terms of function $f_2(\theta;U)$; therefore, we can represent the function $f_2(\theta;U)$ as a linear combination of general and particular solutions:
\begin{widetext}
\begin{equation}
\begin{aligned}
f_2(\theta;U) = P + Q\theta + R\cos(2\theta) + S\sin(2\theta) + E\theta\cos(2\theta) + F\theta\sin(2\theta) + G\theta^2\cos(2\theta) + H\theta^2\sin(2\theta)
\end{aligned}
\label{gen_sol_I}
\end{equation}
\end{widetext}
Here, the coefficients $P, Q, R, S, E, F, G \,\, \text{and} \,\, H$ are functions of the contact angle $\phi$ and the plate speed $U$. We can determine these coefficients by imposing the following boundary conditions:

At the solid surface:
\begin{equation}\label{eq:inertia_BC1}
   -\left.\frac{\partial \psi_c}{\partial r}\right|_{\theta=0}=0 \quad \text { , }\left.\quad \frac{1}{r} \frac{\partial \psi_c}{\partial \theta}\right|_{\theta=0}= 0
\end{equation}

 At the interface:
\begin{equation}\label{eq:inertia_BC2}
   -\left.\frac{\partial \psi_c}{\partial r}\right|_{\theta=\phi}=0 \quad \text { , }\left.\quad\left(-\frac{\partial^2 \psi_c}{\partial r^2}+\frac{1}{r^2} \frac{\partial^2 \psi_c}{\partial \theta^2}+\frac{1}{r} \frac{\partial \psi_c}{\partial r}\right)\right|_{\theta=\phi}=0
\end{equation}

A detailed discussion of the derivation, along with expressions for all coefficients, is provided in Appendix~\ref{apx:inertial-theory}. Equns.~\ref{eqn 3.2} and~\ref{exp_sol_I} denote expressions of streamfunctions in the viscous and weak-inertial regimes, respectively. These models predict streamfunctions in a wedge-shaped flow domain, which cannot be directly compared with the experiments because the interface in the experiments is inherently curved. We modified the model to account for the curvature effect, as discussed in detail in the next section.
\subsection{Modulated wedge solution}\label{sec:wedge_theory}
The HS71 theory and the inertial theory of Varma \textit{et al.} predict the flow dynamics near the contact line for a wedge-shaped flow domain, as shown in Figure~\ref{fig:flat_interface}. In contrast, the experiments involve a curved interface (see figure~\ref{fig:curve_interface}), which restricts a direct comparison with the theoretical models. The study by Chen \textit{et al.} \cite{chen1997velocity} proposed a method to externally incorporate the curved interface into the theoretical model. They showed that by replacing the constant contact angle $\phi$ with a spatially varying angle $\beta$ along the interface, the flow fields corresponding to the curved interface can be determined. The angle $\beta(r)$ is denoted as a modulated wedge angle that varies along the interface, as shown in figure~\ref{fig:curve_interface}.
\begin{figure*}
\centering
    \begin{subfigure}[b]{0.35\textwidth}
        \centering
        \includegraphics[width=\textwidth]{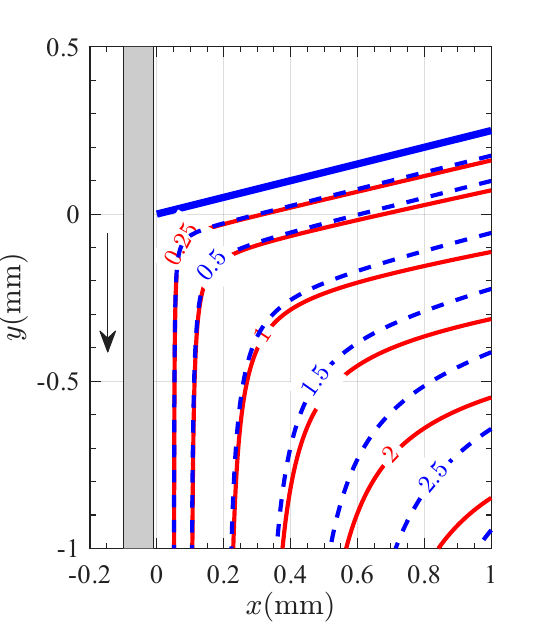}
        \caption{}
        \label{fig:flat_interface_streamfunction}
    \end{subfigure}
    \hspace{5mm}
    \begin{subfigure}[b]{0.35\textwidth}
        \centering
        \includegraphics[width=\textwidth]{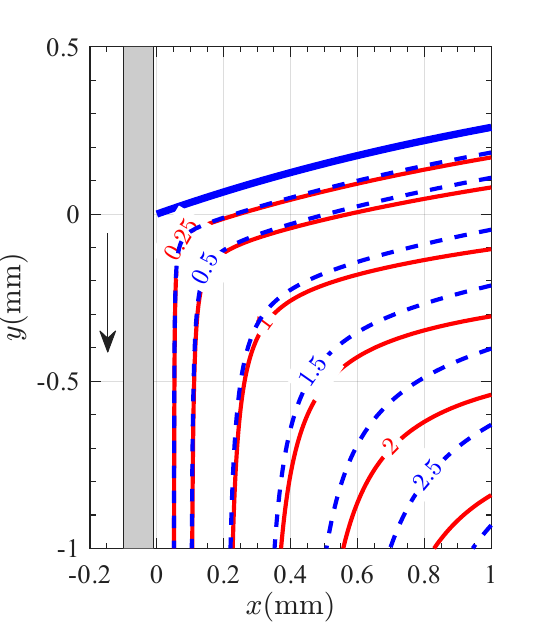}
        \caption{}
        \label{fig:curve_interface_streamfunction}
    \end{subfigure}
\caption{Comparison of streamfunction contours between the flat interface (regression fit) and the curved interface. (a) Comparison between HS71 theory and inertia-corrected theory for the flat interface; linear fit: $y=0.25x$; The flat interface is obtained by performing a linear regression fit ($R^2$ =0.99) to the outlined curved interface. (b) comparison between HS71 theory and inertia-corrected theory for the curved interface; fitting parameters for interface: $c_1$ = 0.40, $c_2$ = 0.09, $c_3$ = -0.41 and $c_4$ = -0.77. The blue dashed curve denotes the streamfunction contour from HS71 theory, and the red solid curve denotes the streamfunction contour from inertia-corrected theory. The blue solid curve denotes the interface (both flat and curved) in the respective plots.}
\label{fig:curve_streamfunction}
\end{figure*}
To obtain the flow fields for the modulated wedge, the modulated wedge angle $\beta(r)$, i.e. the interface shape, must be known a priori. Existing interface shape models \cite{dussan1991,chan2013hydrodynamics,chan2020cox}, such as the Dussan, Rame, and Garoff model (DRG model) \cite{dussan1991}, cannot predict the interface shape independently, as they require a value for a tuning parameter. This parameter can only be determined by comparing the model with the experimental interface shape; therefore, the experimental interface shape is used directly as input to calculate the angle $\beta(r)$. The solution of HS71 theory in a modulated wedge is referred to as viscous-MWS theory. The expression of the streamfunction, $\psi_v$, for the viscous-MWS theory becomes 
\begin{equation}\label{eq:Modulate_wedge_1_p1}
   \psi_v(r,\theta;\beta(r))  = rU \left(\frac{\beta\sin\theta - \theta\sin\beta \cos(\theta -\beta) }{\beta - \sin\beta \cos\beta}\right).
\end{equation}
Similarly, the solution of the inertial theory in a modulated wedge is denoted as the inertial-MWS theory. After replacing the contact angle $\phi$ with $\beta(r)$, the expression of the streamfunction, $\tilde{\psi}$ becomes
\begin{equation}
\label{eqn:int_mws}
    \tilde{\psi}(r,\theta;\beta(r)) = \psi_{v}(r,\theta;\beta(r)) +\psi_{c}(r,\theta;\beta(r))  
\end{equation}
Now equns.~\ref{eq:Modulate_wedge_1_p1} and~\ref{eqn:int_mws} can be used directly to compare with the results obtained from the experiments and simulations. Figures~\ref{fig:flat_interface_streamfunction} and \ref{fig:curve_interface_streamfunction} compare the streamfunction contours predicted by the inertial and inertial-MWS theories with those from the viscous and viscous-MWS theories. Figure~\ref{fig:flat_interface_streamfunction} shows the comparison for the flat interface, whereas Figure~\ref{fig:curve_interface_streamfunction} shows that for the curved interface. Note that curvature did not produce any significant change in the streamfunction contours, while enabling direct comparison with the experimental data. 

\subsection{Interfacial speed}\label{sec:interfacial_speed}
The interfacial speed is one of the key aspects of moving contact lines, defined as the velocity of fluid particles at the interface. According to HS71 theory \cite{huh1971hydrodynamic}, the speed entirely depends on the dynamic contact angle $\phi$. The expression for the interfacial speed for air-liquid systems where air is considered passive (viscosity ratio $\lambda\ll1$) is given as:  
%
\begin{equation}
    v_i^{HS} = u_r|_{\theta=\phi} = U \left(\frac{\phi\cos\phi - \sin\phi}{\phi - \sin\phi \cos\phi}\right),
    \label{eq:Scriven2}
\end{equation}

The HS71 theoretical framework is described in detail in Appendix~\ref{apx:viscous-theory}, while the derivation of the interfacial speed can be found in an earlier study from our group \cite{gupta2024experimental}. The speed $v_i^{HS}$ will be equal to the radial component of the velocity at the interface, i.e. $u_r|_{\theta=\phi}$, which can be calculated directly using the expressions~\ref{eq:vel_streamfunction_relation}, \ref{gensolv} and \ref{coeff_v} given in Appendix~\ref{apx:viscous-theory}. The theoretical framework assumes a flat interface; therefore, the angular component of the velocity at the interface will be equal to zero, i.e., $u_{\theta}|_{\theta=\phi}=0$. To account for the curved interface, we replace the constant wedge angle $\phi$ with a modulated wedge angle $\beta$ in the derivation of the interfacial speed, yielding the following expression:
\begin{equation}\label{eq:interfacial_speed_MWS}
v_i^{\mathrm{MWS}}
= 
\frac{U}{\cos(\alpha - \beta)}\frac{\beta \cos\beta - \sin\beta}
{\left(\beta - \sin\beta \cos\beta\right)} .
\end{equation}
$v_i^{MWS}$ denotes the interfacial speed for a curved interface, where $\alpha$ is the local slope of the interface (see Figure~\ref{fig:curve_interface}). A detailed derivation of the expression is provided in the earlier study from our group \cite{gupta2024experimental}. The derivation of the inertial correction to the interfacial speed is presented in Appendix~\ref{apx:inertial_interfacial_speed}.


\section{Results}\label{sec:results}

This section is organized into two subsections. First, in \S\ref{sec:flowfields}, flow fields are presented in terms of streamfunctions and compared between experimental data  and both the viscous-MWS and inertial-MWS theories over a wide range of Reynolds numbers. Additionally, streamfunction comparisons between simulations and both the viscous and inertial theories are carried out to further support the experimental results.
Finally, in \S \ref{sec:interfacial_speed_results}, the interfacial speeds extracted from the experiments are compared with the viscous-MWS theory.

\subsection{Flow fields}\label{sec:flowfields}
\begin{figure*}[t]
\centering
\resizebox{0.8\textwidth}{!}{ 
\begin{minipage}{\textwidth}
\begin{subfigure}{0.45\textwidth}
  \includegraphics[trim=0mm 0mm 0mm 0mm, clip, width=\textwidth]
  {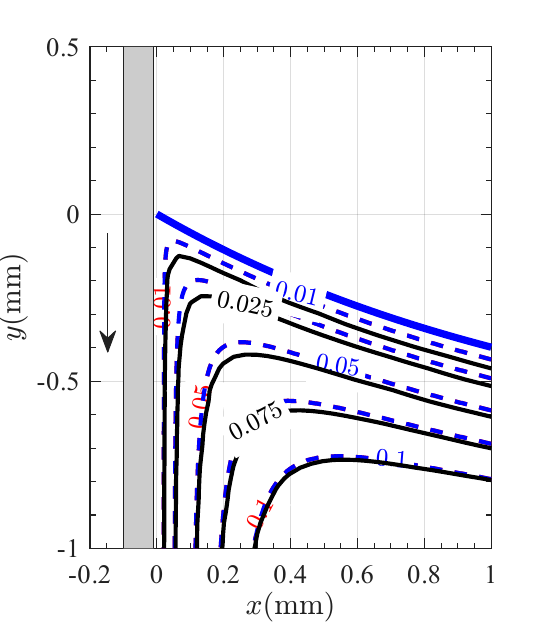}
          \caption{}
        \label{fig:500microns_silicone_oil}
\end{subfigure}
\hfill
\begin{subfigure}{0.45\textwidth}
  \includegraphics[trim=0mm 0mm 0mm 0mm, clip, width=\textwidth]
{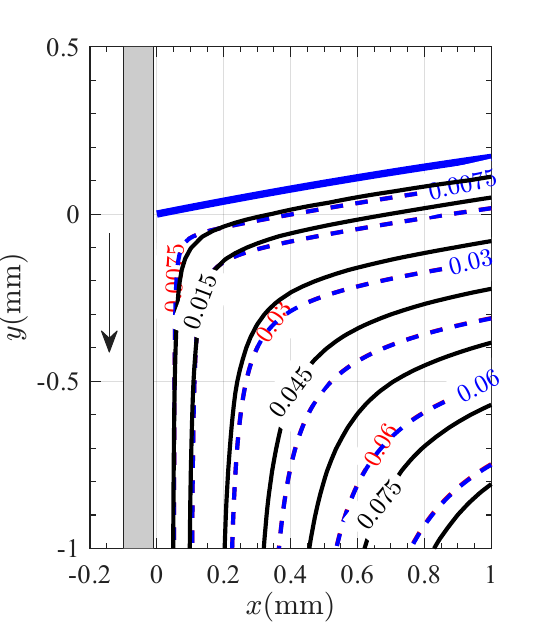}
 \caption{}
\label{fig:150microns_SW50}
\end{subfigure}
\begin{subfigure}{0.45\textwidth}
  \includegraphics[trim=0mm 0mm 0mm 0mm, clip, width=\textwidth]
  {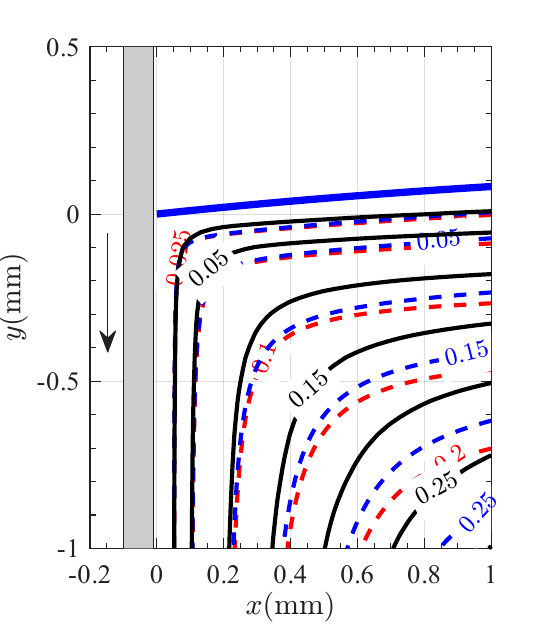}
   \caption{}
\label{fig:500microns_SW48}
  \end{subfigure}
\hfill
\begin{subfigure}{0.45\textwidth}
  \includegraphics[trim=0mm 0mm 0mm 0mm, clip, width=\textwidth]
{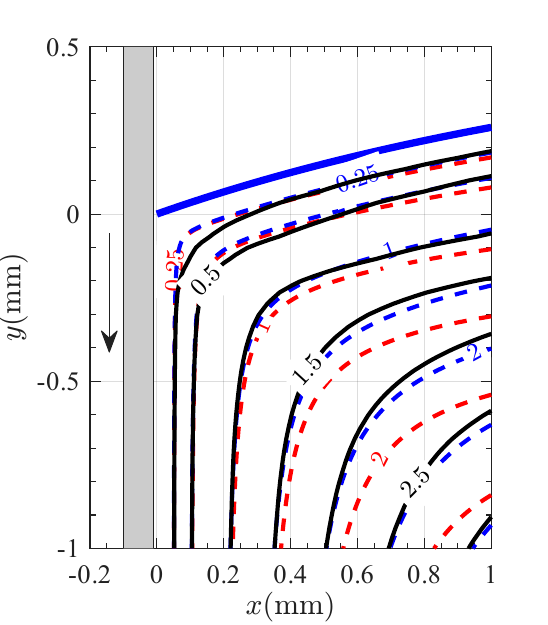}
   \caption{}
\label{fig:5mm_SW40}
\end{subfigure}
\end{minipage}
}
\caption{[Experiments (black solid curve) + viscous theory (blue dashed curve) + inertial correction (red dashed curve)] Contours of the streamfunction obtained from experiments, viscous MWS theory, and inertial MWS theory are superimposed. The grey rectangle represents the solid plate moving downwards, and the blue solid curve represents the interface between air and the liquid phase. Contours are shown for the following Reynolds and capillary numbers: (a) 500 cSt silicone oil at $Re$ = $1.3\times 10^{-3}$ and $Ca$ = $1.4\times 10^{-2}$, fitting parameters for interface: $c_1$ = -0.70, $c_2$ = 0.01, $c_3$ = 0.70 and $c_4$ = -0.81; (b) 50\% sucrose-water at $Re = 3.55 \times 10^{-2}$ and $Ca = 2.6 \times 10^{−5}$, fitting parameters for interface: $c_1$ = 2.66, $c_2$ = -0.19, $c_3$ = -2.39 and $c_4$ = -0.25; (c) 48\% sugar-water at $Re = 1.25 \times 10^{-1}$ and $Ca = 8.16 \times 10^{−5}$, fitting parameters for interface: $c_1$ = 1.64, $c_2$ = -0.22, $c_3$ = -1.64 and $c_4$ = -0.28; (d) 40\% sucrose-water at $Re = 3.02$ and $Ca = 3.3 \times 10^{−4}$, fitting parameters for interface: $c_1$ = 0.40, $c_2$ = 0.09, $c_3$ = -0.41 and $c_4$ = -0.77.}
\label{fig:Inertia_th_viscous_exp}
\end{figure*}
This section examines the flow fields near the moving contact line, which is one of the key results of this study. Flow fields from experiments were obtained using PIV measurements and subsequently used to compute streamfunctions by integrating the continuity equation between adjacent grid points in the $x$–$y$ coordinate system. Streamfunctions predicted by both the viscous-MWS and inertial-MWS theories were systematically compared against the experimentally obtained streamfunctions. The viscous-MWS theory and inertial-MWS theory are discussed in detail in the Appendix~\ref{apx:viscous-theory} and \S\ref{sec:inertial-MWS}, respectively. We compare the streamfunction computed from simulations with the predictions of both viscous-MWS and inertial-MWS theories to support the experimental data. A detailed discussion of the simulations is provided in Appendix~\ref{apx:numerics}. The objective of this section is to demonstrate the effect of inertia on flow configurations and to evaluate how accurately the inertial-MWS theory predicts it.

Figure \ref{fig:Inertia_th_viscous_exp} presents a comparison between experimental streamfunction contours spanning four orders of magnitude in Reynolds number and those predicted by the viscous-MWS and inertial-MWS theories. Figures~\ref{fig:500microns_silicone_oil}, ~\ref{fig:150microns_SW50}, ~\ref{fig:500microns_SW48} and ~\ref{fig:5mm_SW40} correspond to the following air-liquid systems: air-500cSt silicone oil,  air-50\% sucrose water mixture (w/w), air-48\% sugar water mixture (w/w), and air-40\% sucrose water mixture (w/w), as listed in Table \ref{tab:properties_fluids}. The corresponding dimensionless numbers, including the Reynolds number and the Capillary number, are listed in Table~\ref{tab:operating_parameters} for a wide range of plate speeds. 
In all scenarios, fluid particles at the interface move towards the contact line and align with the solid surface; therefore, a rolling motion occurs in the fluid domain. The black solid curves represent the streamfunctions obtained from experiments, the blue dashed curves correspond to the viscous-MWS theory, and the red dashed curves correspond to the inertial-MWS theory. In Figures~\ref{fig:500microns_silicone_oil} and ~\ref{fig:150microns_SW50}, near the contact line, the experimentally obtained streamfunction contours and their corresponding levels agree well with both the viscous-MWS and inertial-MWS theories. As one moves away from the contact line, deviations appear in the contour levels without a significant change in the contour shapes. Such deviations are expected, since viscous effects weaken as one moves away from the contact line. The streamfunctions predicted by the viscous-MWS and inertial-MWS theories overlap, demonstrating that inertial effects remain negligible for Reynolds numbers up to $O(10^{-2})$.

\begin{figure*}[t]
\centering
\resizebox{0.8\textwidth}{!}{ 
\begin{minipage}{\textwidth}
    \begin{subfigure}[b]{0.45\textwidth}
        \centering
        \includegraphics[trim = 0mm 0mm 0mm 0mm, clip, width=\textwidth]{Figures/streamline-plot-comparison-5mm_inertia_dimensional_combined.pdf}
        \caption{}
        \label{fig:5mmps_int}
    \end{subfigure}
    \hfill
    \begin{subfigure}[b]{0.45\textwidth}
        \centering
        \includegraphics[trim = 0mm 0mm 0mm 0mm, clip, width=\textwidth]{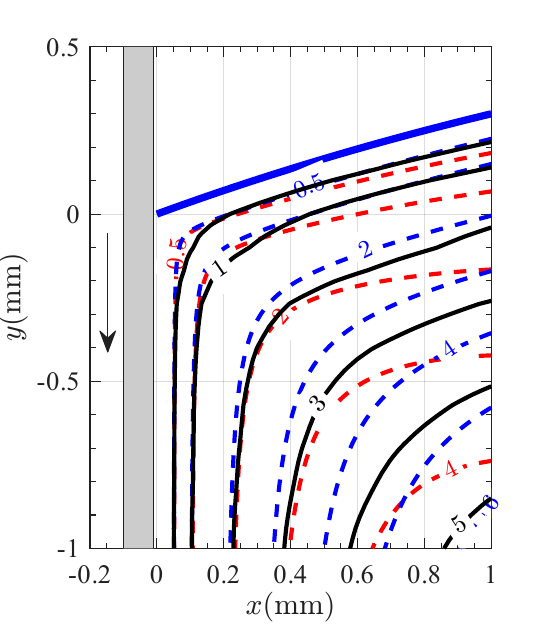}
        \caption{}
        \label{fig:10mmps_int}
    \end{subfigure}
    
    \begin{subfigure}[b]{0.45\textwidth}
         \centering
        \includegraphics[trim = 0mm 0mm 0mm 0mm, clip, width=\textwidth]{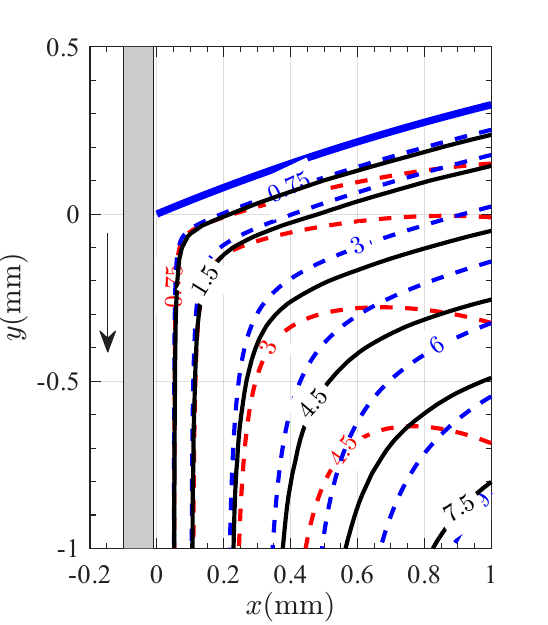}
        \caption{}
        \label{fig:15mmps_int}
    \end{subfigure}
    \hfill
    \begin{subfigure}[b]{0.45\textwidth}
        \centering
        \includegraphics[trim = 0mm 0mm 0mm 0mm, clip, width=\textwidth]{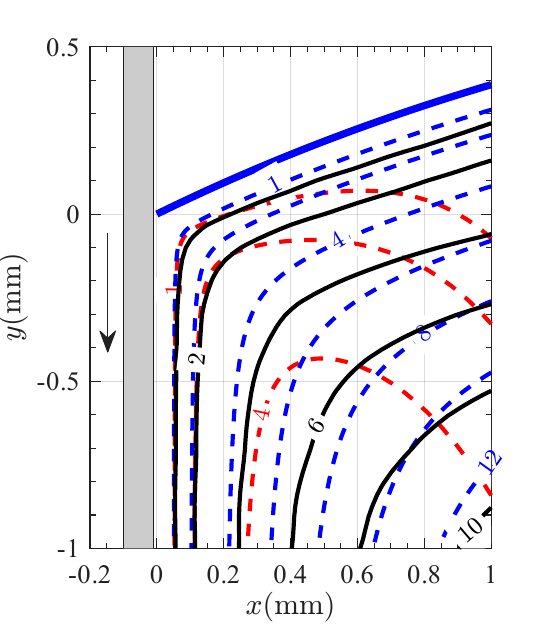}
        \caption{}
        \label{fig:20mmps_int}
    \end{subfigure}
\end{minipage}
}
\caption{[High $Re$ experiments (black solid curve) + viscous theory (blue dashed curve) + inertial correction (red dashed curve)] Contours of the streamfunction obtained from experiments, viscous MWS theory, and inertial MWS theory are superimposed. The grey rectangle represents the solid plate moving downwards, and the blue solid curve represents the interface between air and 40\% sucrose-water mixture. Contours are shown for the following Reynolds and capillary numbers: (a) $Re = 3.02$ and $Ca = 3.3 \times 10^{−4}$, fitting parameters for interface: $c_1$ = 0.40, $c_2$ = 0.09, $c_3$ = -0.41 and $c_4$ = -0.77; (b) $Re = 6.04$ and $Ca = 6.6 \times 10^{−4}$, fitting parameters for interface: $c_1$ = 1.03, $c_2$ = -0.02, $c_3$ = -1.02 and $c_4$ = -0.38; (c) $Re = 9.05$ and $Ca = 9.9 \times 10^{−4}$, fitting parameters for interface: $c_1$ = 0.82, $c_2$ = -0.008, $c_3$ = -0.31 and $c_4$ = -0.56; (d) $Re = 12.1$ and $Ca = 13.3 \times 10^{−3}$, fitting parameters for interface: $c_1$ = 1.15, $c_2$ = -0.05, $c_3$ = -0.54 and $c_4$ = -0.49.}
\label{fig:Inertia_th_intertia_exp}
\end{figure*}
\begin{figure*}[t]
\centering
\resizebox{0.8\textwidth}{!}{ 
\begin{minipage}{\textwidth}
    \begin{subfigure}[b]{0.45\textwidth}
        \centering
        \includegraphics[trim = 0mm 0mm 0mm 0mm, clip, width=\textwidth]{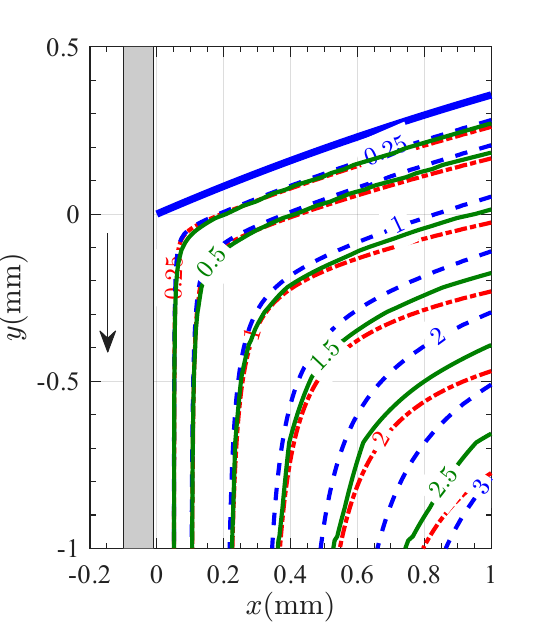}
        \caption{}
        \label{fig:5mmps_int_sims}
    \end{subfigure}
    \hfill
    \begin{subfigure}[b]{0.45\textwidth}
        \centering
        \includegraphics[trim = 0mm 0mm 0mm 0mm, clip, width=\textwidth]{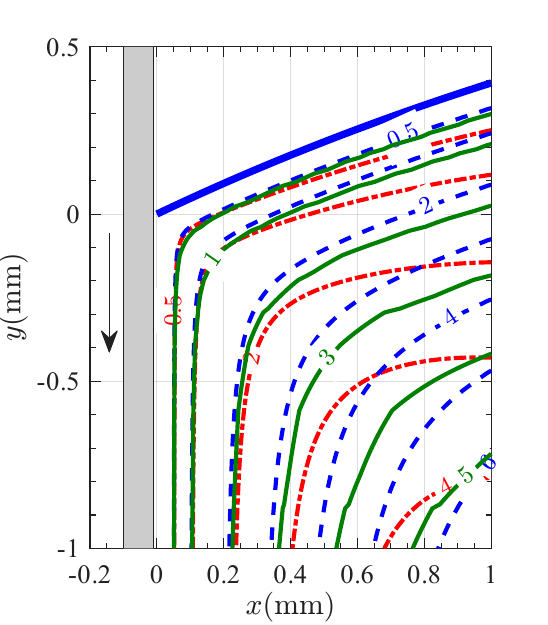}
        \caption{}
        \label{fig:10mmps_int_sims}
    \end{subfigure}
    \begin{subfigure}[b]{0.45\textwidth}
         \centering
        \includegraphics[trim = 0mm 0mm 0mm 0mm, clip, width=\textwidth]{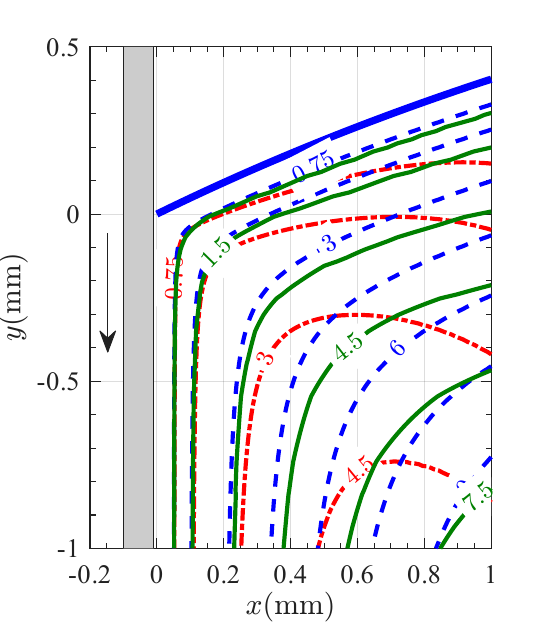}
        \caption{}
        \label{fig:15mmps_int_sims}
    \end{subfigure}
    \hfill
    \begin{subfigure}[b]{0.45\textwidth}
        \centering
        \includegraphics[trim = 0mm 0mm 0mm 0mm, clip, width=\textwidth]{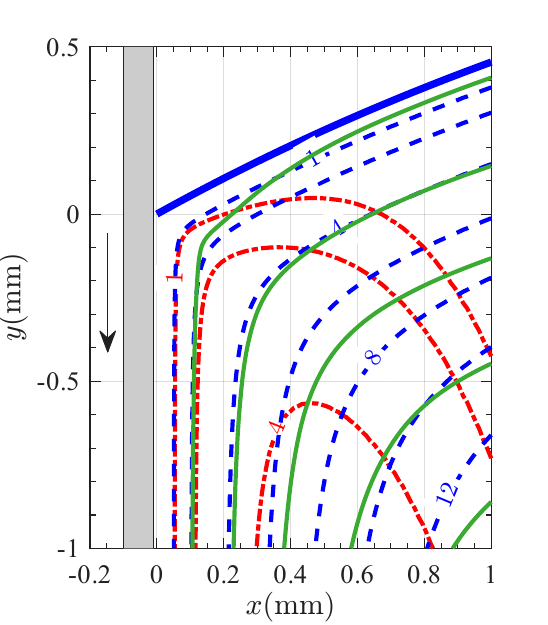}
        \caption{}
        \label{fig:20mmps_int_sims}
    \end{subfigure}
\end{minipage}
}
\caption{[Simulations (green solid curve) + viscous theory (blue dashed curve) + inertial correction (red dash-dot curve)] Contours of the streamfunction obtained from simulations, viscous MWS theory, and inertial MWS theory are superimposed. The grey rectangle represents the solid plate moving downwards, and the blue solid curve represents the interface between air and 40\% sucrose-water mixture. Contours are shown for the following Reynolds and capillary numbers: (a) $Re = 3.02$ and $Ca = 3.3 \times 10^{−4}$  (b) $Re = 6.04$ and $Ca = 6.6 \times 10^{−4}$ (c) $Re = 9.05$ and $Ca = 9.9 \times 10^{−4}$ (d) $Re = 12.1$ and $Ca = 13.3 \times 10^{−3}$}
\label{fig:Sims-Inertia}
\end{figure*}

In Figures~\ref{fig:500microns_SW48} and~\ref{fig:5mm_SW40}, the deviations between the streamfunctions predicted by the viscous-MWS theory and the inertial-MWS theory are non-trivial. In particular, the streamfunction predicted by the inertial-MWS theory is deflected away from the interface compared to that predicted by the viscous-MWS theory. These figures correspond to Reynolds numbers in the range of $10^{-1}$ to 1. The streamfunction obtained from the experiments also deflects in the same direction as predicted by the inertial-MWS theory, as shown in Figures~\ref{fig:500microns_SW48} and~\ref{fig:5mm_SW40}. These deflections occur primarily away from the contact line, while the inertial-MWS predictions and experimental data show good agreement with the viscous-MWS predictions near the contact line. In other words, the deflections between streamfunctions are smaller near the contact line and progressively increase moving away from it. Figure~\ref{fig:5mm_SW40} shows larger deflections than Figure~\ref{fig:500microns_SW48} because the Reynolds number is higher in the former case.

\begin{figure*}[t]
\centering
\resizebox{0.8\textwidth}{!}{ 
\begin{minipage}{\textwidth}
    \begin{subfigure}[b]{0.45\textwidth}
         \centering
        \includegraphics[trim = 0mm 0mm 0mm 0mm, clip, width=\textwidth]{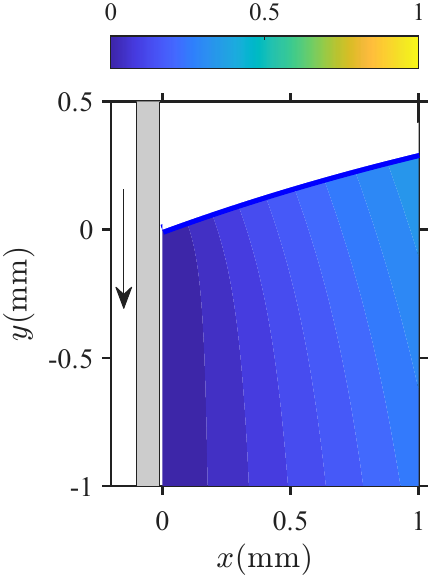}
        \caption{}
        \label{fig:streamfunction_spatial_variation_Re_6}
    \end{subfigure}
    \hspace{5mm}
    \begin{subfigure}[b]{0.45\textwidth}
         \centering
        \includegraphics[trim = 0mm 0mm 0mm 0mm, clip, width=\textwidth]{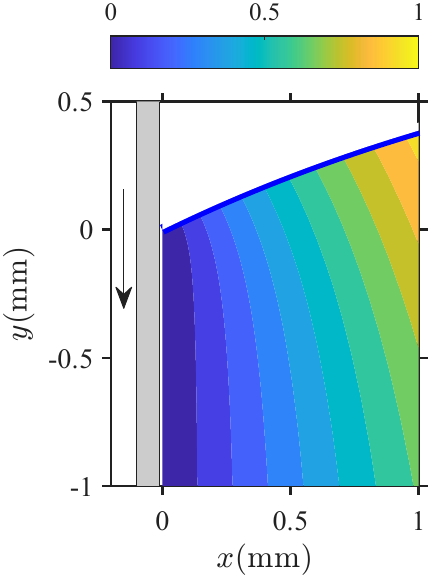}
        \caption{}
        \label{fig:streamfunction_spatial_variation_Re_12}
    \end{subfigure}
\end{minipage}
}
\caption{ The spatial variation of $\psi_c/\psi_{\nu}$ is shown for two different Reynolds number. $\psi_c$ represents the inertial correction to the viscous streamfunction $\psi_{\nu}$. (a) $Re = 6.04$ and $Ca = 6.6 \times 10^{−4}$ (b) $Re = 12.1$ and $Ca = 13.3 \times 10^{−3}$.}
\label{fig:streamfunction_spatial_variation}
\end{figure*}
Figure~\ref{fig:Inertia_th_intertia_exp} shows a comparison of streamfunction contours between experiments and both the viscous-MWS and inertial-MWS theories for Reynolds numbers greater than one. The streamfunctions obtained from experiments correspond to the air–40\% sucrose water system. Using the same notation as in Figure~\ref{fig:Inertia_th_viscous_exp}, Figures~\ref{fig:5mmps_int} and~\ref{fig:10mmps_int} show reasonable deflections between the streamfunction contours predicted by the viscous-MWS theory and those from the inertial-MWS theory. These deflections become pronounced at higher Reynolds numbers, as shown in Figures~\ref{fig:15mmps_int} and~\ref{fig:20mmps_int}. These figures correspond to Reynolds numbers on the order of $O(10)$. The streamfunctions obtained from the experiments, although showing deflections in the same direction away from the contact line, do not exhibit the same magnitude of deflection as predicted by the inertial-MWS theory. This limits the predictive capability of the inertial-MWS theory, as the deflections become unreasonable compared to the experimental data.




Figure~\ref{fig:Sims-Inertia} shows a comparison between streamfunctions calculated from simulations and streamfunctions predicted by both viscous-MWS and inertial-MWS theories. These simulations are performed at the same Reynolds numbers as those presented in Figure~\ref{fig:Inertia_th_intertia_exp}. We replicated Figure~\ref{fig:Inertia_th_intertia_exp} primarily to support the experimental evidence through numerical simulations. Figures~\ref{fig:5mmps_int_sims}, \ref{fig:10mmps_int_sims}, \ref{fig:15mmps_int_sims}, and \ref{fig:20mmps_int_sims} exhibit deflections similar to those observed in the experiments shown in Figure~\ref{fig:Inertia_th_intertia_exp}. The predictions from the viscous-MWS and inertial-MWS theories, calculated using the simulation interface shapes as input, provide similar results, as shown in Figure~\ref{fig:Inertia_th_intertia_exp}. This confirms that the inertial-MWS theory yields reasonable predictions only within a narrow range of Reynolds numbers. 

We further investigated the limitations of the inertial-MWS theory by analyzing the spatial variation of the streamfunctions ratio $\psi_c/\psi_{\nu}$. Here, $\psi_c$ denotes the streamfunction corresponding to the inertial correction to the viscous streamfunction $\psi_{\nu}$. The formulation remains valid only when $\psi_{\nu} \gg \psi_c$, i.e., $\psi_c/\psi_{\nu} \ll 1$. In principle, the ratio $\psi_c/\psi_{\nu}$ remains small ($\psi_c/\psi_{\nu} \ll 0.1$) near the contact line and increases progressively with distance from the contact line. Figure~\ref{fig:streamfunction_spatial_variation} shows the spatial variation of the streamfunction ratio in the flow domain for two different Reynolds numbers. In Figure~\ref{fig:streamfunction_spatial_variation_Re_6}, the maximum value of the streamfunction ratio is approximately 0.3 across the spatial domain. The corresponding streamfunction comparison between the viscous-MWS and inertial-MWS theories shows good agreement, as shown in Figure~\ref{fig:10mmps_int}. This comparison corresponds to the Reynolds number of $Re = 6.04$. The streamfunction ratio $\psi_c/\psi_{\nu}$ for $Re = 12.1$ exceeds 0.5 away from the contact line across much of the spatial domain, as illustrated in Figure~\ref{fig:streamfunction_spatial_variation_Re_12}. This large value indicates that the present formulation may no longer be valid. In other words, higher-order terms in the inertial perturbation expansion become important; therefore, the validity of the present formulation is limited. This is consistent with the pronounced deviation of the inertial-MWS streamfunction shown in Figure~\ref{fig:20mmps_int}.  







\subsection{\label{sec:interfacial_speed_results}Interfacial speed}
\begin{figure*}[t]
 \centering
 \includegraphics[trim = 0mm 0mm 0mm 0mm, clip, width=0.7\textwidth]{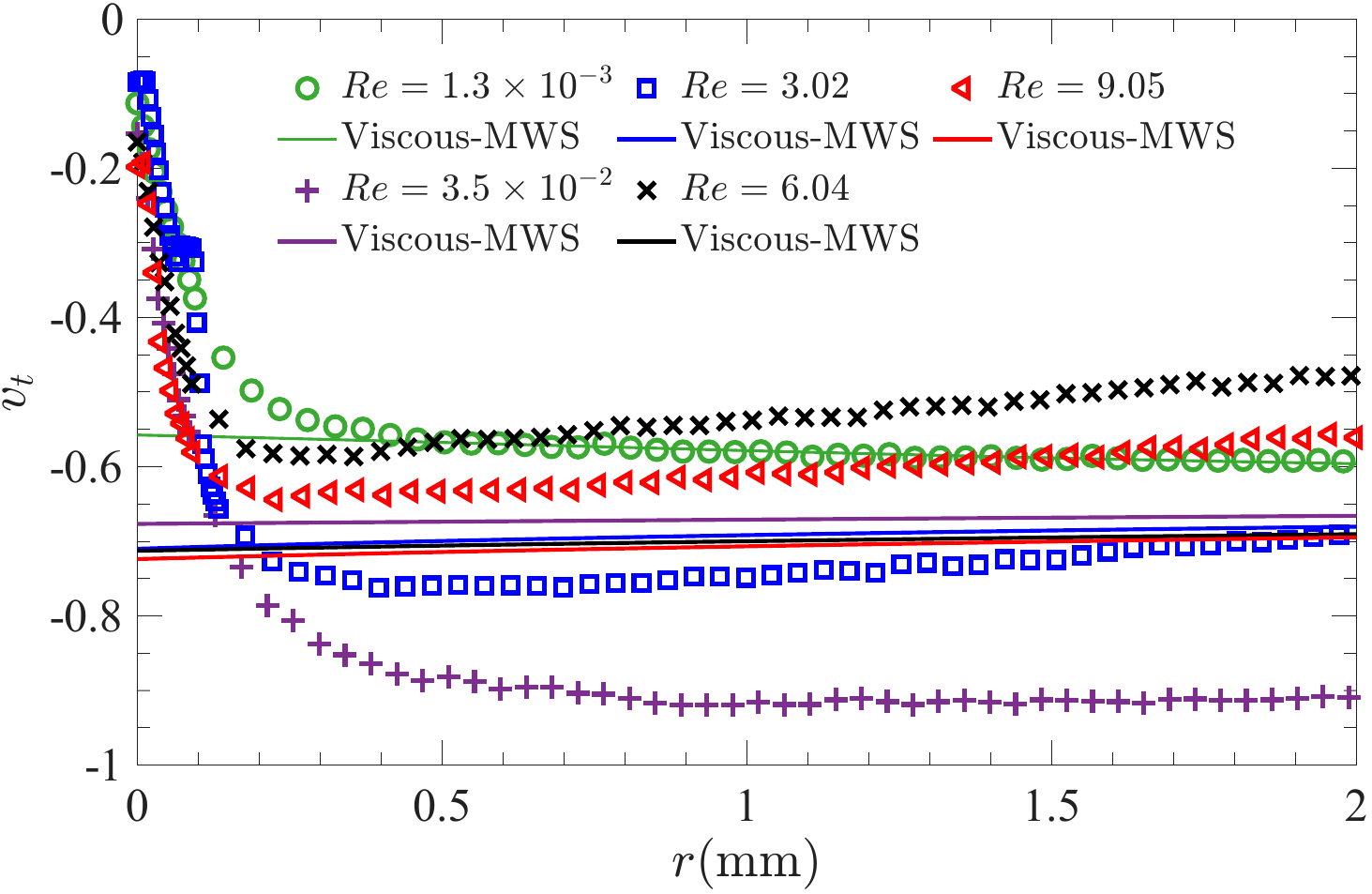}
\caption{The variation of the tangential velocity along the interface for different fluid-fluid systems as a function of the radial position from the contact line. The tangential velocity is presented as the non-dimensional velocity ($v_t$), normalized by the plate speed ($U$). The velocities are extracted at a location 50~$\mu$m from the interface to avoid fluctuations that occur near the interface, particularly at high plate speeds. The velocities are shown for the following Reynolds and capillary numbers: \textcolor[rgb]{0.23,0.67,0.20}{\Large$\boldsymbol{\circ}$}: $Re = 1.3 \times 10^{-3}$ and $Ca = 1.4 \times 10^{−2}$, $\theta_d$ = 60$^\circ$; \textcolor[rgb]{0.5, 0.0, 0.5}{\Large$\boldsymbol{+}$}: $Re = 3.5 \times 10^{-2}$ and $Ca = 2.6 \times 10^{−5}$, $\theta_d$ = 101.2$^\circ$; $\color{blue}\large\boldsymbol{\square}$: $Re = 3.02$ and $Ca = 3.3 \times 10^{−4}$, $\theta_d$ = 109.5$^\circ$; $\color{black}\Large\boldsymbol{\times}$: $Re = 6.04$ and $Ca = 6.6 \times 10^{−4}$, $\theta_d$ = 110.1$^\circ$; $\color{red}\large\boldsymbol{\triangleleft}$: $Re = 9.05$ and $Ca = 9.9 \times 10^{−4}$, $\theta_d$ = 112.7$^\circ$.}
\label{fig:Interfacial_speed_tan}
\end{figure*}

%
The interfacial speed is defined as the velocity of fluid particles along the interface. The direction of this velocity determines the flow configuration within the domain, making it a key aspect of moving contact lines. In the experiments, the interfacial speed is extracted by projecting the flow fields obtained from the PIV measurements to a location 50~$\mu$m from the interface and then rigorously compared with predictions from theoretical models. The theoretical interfacial speed is computed using the HS71 and viscous-MWS theories presented in \S\ref{sec:interfacial_speed}.

Figure~\ref{fig:Interfacial_speed_tan} shows the interfacial speed for four different Reynolds numbers, ranging from $O(10^{-3})$ to $O(10)$. The interfacial speed is non-dimensionalized by the plate speed and its variation along the interface is examined as a function of the distance $r$ from the contact line. The speed is negative for all cases shown in Figure~\ref{fig:Interfacial_speed_tan}, indicating that the fluid particles at the interface are moving toward the contact line; consequently, a rolling motion is expected to occur in the flow domain. This inference, drawn solely from the direction of the interfacial speed, is consistent with the flow configurations reported in \S\ref{sec:flowfields}. The experimental interfacial speed along the interface exhibits two distinct trends: one in the near-field region and the other in the far-field region. The experimental interfacial speed at low Reynolds numbers shows a nearly constant trend in the far field as shown with a marker `$\textcolor[rgb]{0.23,0.67,0.20}{\text{o}}$' in Figure~\ref{fig:Interfacial_speed_tan}. The speed shows a good agreement with the interfacial speed predicted by viscous-MWS theory. The speed predicted by viscous-MWS theory accounts for curvature effects of the interface, leading to a slight increase (decrease) relative to the HS71 prediction for acute (obtuse) contact angles. This speed corresponds to the air-silicone oil system, whereas the other speeds correspond to air-water-based systems. The interfacial speed corresponding to the `${\textcolor{myPurple}{\boldsymbol{+}}}$' marker also exhibits a constant trend in the far field. However, its magnitude is higher than that of the viscous-MWS prediction, which may be due to the effect of the hydrophobic coating on the plate. For experiments at Reynolds numbers of $O(1)$, the interfacial speed exhibits a monotonic decay with an approximately constant slope. This decrease is consistently observed for all high Reynolds number cases shown in Figure~\ref{fig:Interfacial_speed_tan}, suggesting the influence of inertial effects on the interfacial speed. This conclusion is consistent with the discussion of inertial effects on interfacial speed presented in Figure~\ref{fig:interfacial_viscous_inertia} in the Appendix~\ref{apx:inertial_interfacial_speed}. Theoretical analysis further demonstrates that inertia leads to a monotonic decrease in the interfacial speed. The magnitudes of the speeds are not consistent with the viscous-MWS predictions, which may be due to a combination of inertia and the hydrophobic coating on the plate.

In the near-field region, the interfacial speed exhibits a sharp decrease as the contact line is approached. This trend persists over a wide range of Reynolds numbers, indicating that the interfacial speed is largely independent of inertial effects. The rapid decrease suggests that fluid particles at the interface undergo a strong deceleration near the contact line and reach it with a very low velocity while aligning with the solid surface. As they align with the solid, the particles gradually accelerate to match the solid velocity, suggesting the presence of a slip region along the wall.

The interfacial speed in the viscous regime (curves with markers `$\textcolor[rgb]{0.23,0.67,0.20}{\text{o}}$' and `${\textcolor{myPurple}{\boldsymbol{+}}}$') exhibits two distinct trends: a sharp decrease near the contact line and a constant speed in the far field. These trends correspond to the microscopic and viscous regimes, respectively. The interfacial speed corresponding to the marker `$\textcolor{blue}{\boldsymbol{\square}}$' in Figure~\ref{fig:Interfacial_speed_tan} clearly exhibits three distinct trends: a sharp decrease near the contact line, a monotonic decay away from the contact line, and a constant speed connecting the two. The monotonic decay in the inertial regime progressively diminishes the extent of the constant viscous regime in the far-field. The interfacial speeds at higher Reynolds numbers (curves with markers `$\textcolor[rgb]{0,0,0}{\boldsymbol{\times}}$' and `$\textcolor[rgb]{1,0,0}{\triangleleft}$') show that the constant region becomes progressively smaller, while the effect of inertia on the speed increases.  

\section{Summary and Discussions}\label{sec:conclusion}
This study examined the inertial effects on the flow dynamics in the vicinity of the moving contact line. A combination of experiments, simulations, and rigorous testing of theoretical models was carried out. The experiments were conducted using the plate immersion configuration over a wide range of Reynolds numbers from $O(10^{-3})$ to $O(10)$. The capillary numbers span from $O(10^{-5})$ to $O(10^{-2})$. This wide range of Reynolds numbers was selected to clearly observe the transition from the viscous-dominated regime to the inertia-dominated regime. Theoretical models, including the viscous-MWS and inertial-MWS models, were used to carry out streamfunction comparisons with the experimental data. We used these comparisons to determine the effect of inertia on the flow dynamics near the moving contact line and to evaluate the predictive capability of the inertial–MWS theory in capturing inertial effects.

We presented experimental results obtained using air-liquid systems such as air-500 cSt silicone oil, air-50\% sucrose water mixture, air-48\% sugar water mixture and air-40\% sucrose water mixture. These different fluid phases, ranging from air-500 cSt silicone oil with a Reynolds number of $10^{-3}$ to air-40\% sucrose water mixtures with Reynolds numbers of $O(10)$, allowed us to systematically probe and identify inertial effects.
For all experiments, we recorded the motion of the seeding particles within a $4.2\text{mm} \times 4.2\text{mm}$ region near the moving contact line in the two-dimensional plane using high-speed imaging and particle image velocimetry. The recorded images were then post-processed using the in-house PIV software and MATLAB to obtain flow fields, interface shapes, and interfacial speeds. The experimental setup, along with the design methodology, is detailed in \S\ref{sec:experimental_methodology}.

We rigorously compared the streamfunction contours, directly derived from the experimentally obtained flow fields, with the predictions of the viscous-MWS and inertial-MWS theories. The inertial theory \cite{varma2021inertial} extends the Huh \& Scriven viscous model \cite{huh1971hydrodynamic} by incorporating the leading-order inertial correction via the perturbation expansion in the local Reynolds number, $\varrho$. Both viscous and inertial theories assumed a flat interface in their formulation, leading to inconsistencies in direct comparison with experimental results. To account for the curved interface, we followed a simple procedure, described in \S \ref{sec:wedge_theory} \cite{gupta2023}, in which the experimental interface is used directly as input to the theories. A detailed discussion of viscous-MWS and inertial-MWS theories is provided in Appendix~\ref{apx:viscous-theory}, \S\ref{sec:inertial-MWS}, Appendix~\ref{apx:inertial-theory} and \S\ref{sec:wedge_theory}.

The present study compared the streamfunction contours for fluid phases between the experimental data and the predictions of the viscous-MWS and inertial-MWS theories. This comparison spans Reynolds numbers from $O(10^{-3})$ (Figure~\ref{fig:500microns_silicone_oil}) to $O(10)$ (Figure~\ref{fig:20mmps_int}). Streamfunction contours from viscous-MWS and inertial-MWS theories completely overlap for Reynolds numbers up to $O(10^{-2})$, as illustrated in Figures~\ref{fig:500microns_silicone_oil} and \ref{fig:150microns_SW50}. This indicates that the comparisons shown in Figures~\ref{fig:500microns_silicone_oil} and \ref{fig:150microns_SW50} are purely in the viscous regime, with the theories showing good agreement with the experimental data. Further ahead, for Reynolds number greater than $O(10^{-2})$, as shown in Figures~\ref{fig:500microns_SW48} and \ref{fig:5mm_SW40}, the streamfunction contours reveal deviations between the viscous-MWS and inertial-MWS theories.
These comparisons confirm that inertia does not alter the flow configuration near the contact line, but rather deflects the streamfunction contours away from the interface. These deflections progressively increase with distance from the contact line and with increasing Reynolds number. For Reynolds numbers greater than $O(1)$, the inertial-MWS theory begins to predict unexpected large deflections in the streamfunction contours, as shown in Figure~\ref{fig:Inertia_th_intertia_exp}. These large deflections occur primarily away from the contact line, while good agreement is observed in the streamfunction contours near the contact line. In contrast, the experimental streamfunction contours exhibit reasonable deflections away from the contact line even at high Reynolds numbers. The simulation results, as shown in Figure~\ref{fig:Sims-Inertia}, demonstrate trends similar to those observed in the experiments. Moreover, the streamfunction associated with the inertial correction is compared with the viscous streamfunction using a ratio $\psi_c/\psi_{\nu}$ to examine the applicability of the inertial correction, as illustrated in Figure~\ref{fig:streamfunction_spatial_variation}. The correction is expected to provide accurate results when this ratio is very small ($\psi_c/\psi_{\nu} \ll 1$). Figure~\ref{fig:streamfunction_spatial_variation_Re_12} shows that the ratio exceeds $0.5$ away from the contact line, highlighting the limitation of the inertial correction. In summary, these comparisons demonstrate that the inertial-MWS theory provides reasonable predictions only within a narrow range of Reynolds numbers, from $O(10^{-1})$ to $O(1)$.

We compared the interfacial speeds from the experiments with predictions from the viscous-MWS theory. Figure~\ref{fig:Interfacial_speed_tan} shows that the speed along the interface for all experiments is negative, indicating that the fluid particles are moving toward the contact line; therefore, the rolling motion is expected. This inference from the interfacial speed is consistent with the streamfunction contours shown in Figure~\ref{fig:Inertia_th_viscous_exp}, which also shows the rolling motion in the flow domain. In the viscous regime, the interfacial speed remains nearly constant away from the contact line, whereas it decays monotonically with an approximately constant slope in the inertial regime. This monotonic decay of the interfacial speed is consistent with the prediction of the inertial-MWS theory, as illustrated in Figure~\ref{fig:interfacial_viscous_inertia} and discussed in Appendix~\ref{apx:inertial_interfacial_speed}. The interfacial speed near the contact line exhibits a nearly identical trend over a wide range of Reynolds numbers, indicating that it is largely independent of inertial effects. The interfacial speed (curve with marker `$\textcolor{blue}{\boldsymbol{\square}}$' in Figure~\ref{fig:Interfacial_speed_tan}) clearly exhibits all three distinct trends: a sharp decrease near the contact line, a monotonic decay away from the contact line, and a constant speed connecting the near- and far-field regions. This constant trend of speed becomes progressively smaller as the Reynolds number increases.

In summary, inertia does not alter the flow configuration near the contact line; rather, it only causes deflections in the streamfunction contours away from the interface. The inertial-MWS theory predicts this effect well only within a narrow range of Reynolds numbers from $O(10^{-1})$ to $O(1)$. For high $Re$, the theory predicts large deflections that are not consistent with the experimental results, whereas for low $Re$, the predictions coincide with the viscous-MWS theory, indicating negligible inertial effects. The interfacial speeds show a gradual decrease away from the contact line due to inertia, whereas they are expected to remain nearly constant in the viscous regime. In the near field, a similar trend of sharp decrease in speed is observed, irrespective of the Reynolds number. This deceleration of the interfacial speed near the contact line, observed across a wide range of Reynolds numbers, is consistent with a universal mechanism for resolving the contact line singularity.


\vspace{5mm}
\begin{acknowledgments}
We acknowledge the support of the Anusandhan National Research Foundation (formerly Science and Engineering Research Board), Department of Science and Technology, Government of India, for funding this research through grant no. CRG/2021/007096. VSAS thanks the Prime Minister's Research Fellowship for financial support.
\end{acknowledgments}

\vspace{5mm}
\noindent \textbf{DATA AVAILABILITY}\\
The data that support the findings of this study are available from the corresponding author upon reasonable request.

\bigskip

\appendix

\section{Viscous theory} \label{apx:viscous-theory}
Huh \& Scriven \cite{huh1971hydrodynamic} reported that the biharmonic equation (see equn.~\ref{eqn 3.1}) governs the dynamics near the wedge, as shown in figure~\ref{fig:flat_interface}. The streamfunction is related to the velocity components with the following expression:
\begin{equation}\label{eq:vel_streamfunction_relation}
    u_r = \dfrac{1}{r}\dfrac{\partial \psi}{\partial \theta}, \quad u_{\theta} =- \dfrac{\partial \psi}{\partial r}. 
\end{equation}
Equn.~\ref{eqn 3.1} is a fourth-order linear PDE whose general solution can be expressed as a linear combination of linearly independent solutions. A commonly accepted form of the solution is given by:
\begin{equation}
\label{eqn 3.2_1}
    \psi_{v}(r,\theta) = r f_1(\theta;U),
\end{equation}
where $f_1(\theta;U)$ is a dimensional function with units of velocity and depends only on the wedge angle and the plate speed. By substituting equn.~\ref{eqn 3.2_1} in the governing biharmonic equn.~\ref{eqn 3.1}, we obtain
\begin{equation}
\label{gensolvf_f1}
    f_1 + 2 f''_{1} + f''''_{1}= 0,
\end{equation}
where $()'$ denotes differentiation with angle $\theta$. The solution of the above equation reduces to the form
\begin{equation}
\label{gensolv}
    f_1(\theta;U) = A\cos\theta + B\sin\theta  + C  \theta \cos\theta + D \theta \sin\theta,
\end{equation}
where the dimensional constants $A, B, C$ and $D$ are determined by imposing relevant boundary conditions on the moving wall and the fluid-gas interface. In the present study, the fluid phase A is always a gas; therefore, it is assumed to be passive.

At the solid surface, applying the no-penetration and no-slip conditions, we obtain
\begin{equation}\label{eq:viscous_BC1}
   -\left.\frac{\partial \psi_v}{\partial r}\right|_{\theta=0}=0, \quad \text {}\left.\quad \frac{1}{r} \frac{\partial \psi_v}{\partial \theta}\right|_{\theta=0}= U,
\end{equation}
where $U$ is the velocity of the plate. Similarly, applying the no-penetration and stress-free boundary conditions at the interface, we obtain
\begin{align}\label{eq:viscous_BC2}
   -\left.\frac{\partial \psi_v}{\partial r}\right|_{\theta=\phi}=0, \quad \text { }\left.\quad\left(-\frac{\partial^2 \psi_v}{\partial r^2}+\frac{1}{r^2} \frac{\partial^2 \psi_v}{\partial \theta^2}+\frac{1}{r} \frac{\partial \psi_v}{\partial r}\right)\right|_{\theta=\phi}=0.
\end{align}
The coefficients $A$, $B$, $C$ and $D$ in equn.~\ref{gensolv} can easily be obtained from the four boundary conditions~\ref{eq:viscous_BC1} and \ref{eq:viscous_BC2}, which yield:
\begin{equation} 
\begin{aligned}
        A = \;& 0; \qquad 
    B = -\frac{U\phi}{\phi - \sin\phi\cos\phi}; \qquad
    C = \frac{U\sin\phi\cos\phi} {\phi - \sin\phi\cos\phi}; \qquad \\  
    D = & \frac{U\sin^2 \phi}{\phi - \sin\phi\cos\phi}.
\end{aligned}
    \label{coeff_v}
\end{equation}

\section{Inertial theory} \label{apx:inertial-theory}
The theory comprises the HS71 viscous model together with the leading-order inertial term, derived via a perturbation expansion in the local Reynolds number. Substituting the streamfunction given in equn.~\ref{exp_sol_I} into the governing equn.~\ref{gov_eqn_I} yields:
\begin{equation}
\label{gensolvf_1}
f_2'''' + 4 f_2'' = -2 f_1 f_1' - f_1' f_1'' - f_1 f_1''',
\end{equation}
and we define
\[
F_2(\theta;U) := f_2'''' + 4 f_2'', \quad
F_1(\theta;U) := -2 f_1 f_1' - f_1' f_1'' - f_1 f_1'''.
\]
The expressions for $f_1({\theta;U})$ and $f_2(\theta;U)$ are given in equns.~\ref{gensolv} and \ref{gen_sol_I}, respectively. After substituting these expressions into equn.~\ref{gensolvf_1}, we obtain
\begin{widetext}
\begin{align}
    F_1(\theta;U) &= (2BC + C^2 - D^2)\sin2\theta - 2(B+C)D\cos2\theta 
    + 4CD\theta \sin2\theta + 2(C^2 - D^2)\theta\cos2\theta, \label{F1_theta} \\
    F_2(\theta;U) &= (16E - 40H)\sin2\theta - (16F + 40G)\cos2\theta 
    + (32G)\theta \sin2\theta - (32H)\theta\cos2\theta. \label{F2_theta}
\end{align}
\end{widetext}
The function $f_2(\theta;U)$ contains eight coefficients, four ($E, F, G, H$) of which are determined either from equn.~\ref{gensolvf_1} or by equating equns.~\ref{F1_theta} and \ref{F2_theta}. The remaining four coefficients ($P, Q, R, S$) are determined using the boundary conditions \ref{eq:inertia_BC1} and \ref{eq:inertia_BC2}, which yield:
\begin{widetext}
\begin{equation}
\begin{aligned}
\,& P + R = 0,   \\
& Q + 2S + E = 0, \\
& P + Q\phi + (R + E\phi + G\phi^2)\cos2\phi+ (S + F\phi + H\phi^2)\sin2\phi  = 0,  \\
& \cos2\phi(-4R + 4F + 2G - 4E\phi + 8H\phi - 4G\phi^2) + \sin2\phi(-4S - 4E + 2H - 4F\phi -4H\phi^2 - 8G\phi) = 0.
\end{aligned}
\label{Common_coff}
\end{equation}
\end{widetext}
We solve equns.~\ref{gensolvf_1} and \ref{Common_coff} to obtain the following expressions for $P, Q, R, S , E, F, G \,\, \text{and} \, \, H$:
\begin{widetext}
\begin{equation*}  
\begin{aligned}
P &= -\phi \, Q, \\
Q &= \frac{\sin\phi \left(4 \phi \left(4 \phi \cos\phi + 2 \sin\phi - \sin 3\phi\right)- 3 \sin\phi \sin 4\phi\right)}
{16 \left(2 \phi \cos 2\phi - \sin 2\phi\right) \left(\sin 2\phi -2 \phi \right)^2}, \\
R &= \phi \, Q, \\
S &= -\frac{C D}{16} \left (\frac{\phi (4 \phi + 3 \sin 2\phi)}{2 \phi \cos 2\phi - \sin 2\phi} \right)
\end{aligned}
\end{equation*}

\vspace{1em} 

\begin{equation} \label{inertial_coeff}
\begin{aligned}
E &= \frac{4 B C + 3 (D^2 - C^2)}{32}, \qquad \qquad \qquad \qquad \qquad \qquad \qquad \\
F &= \frac{D(2 B - 3 C)}{16}, \\
G &= \frac{C D}{8}, \\
H &= \frac{D^2-C^2}{16}
\end{aligned}
\end{equation}
\end{widetext}
The coefficients depend on the contact angle $\phi$ and the plate speed $U$, since they are determined by the coefficients $B, C,$ and $D$. 
\section{Expression for inertial interfacial speed}\label{apx:inertial_interfacial_speed}
The leading inertial correction to the interfacial speed can be calculated using the equns.~\ref{eq:vel_streamfunction_relation} in Appendix~\ref{apx:viscous-theory} and~\ref{exp_sol_I} in \S\ref{sec:inertial-MWS}. Similar to the HS71 model, the speed will be denoted by $u_r|_{\theta=\phi}$ and given as
\begin{equation}
    u_r|_{\theta=\phi} = f_1^{'} + \frac{r}{\nu}f_2^{'}.
    \label{eq:Scriven_inertia}
\end{equation}
Here, $f_1^{'}(\theta;U) = v_i^{HS}$ and $f_2^{'}(\theta;U)$ can be calculated using equns.~\ref{gen_sol_I} and~\ref{inertial_coeff}. Note that the expression varies linearly with $r$, in contrast to the HS71 model, which predicts the interfacial speed to be independent of the radial distance from the contact line. 
\begin{figure}
    \centering
        \includegraphics[trim = 0mm 0mm 0mm 0mm, clip, width=0.5\textwidth]{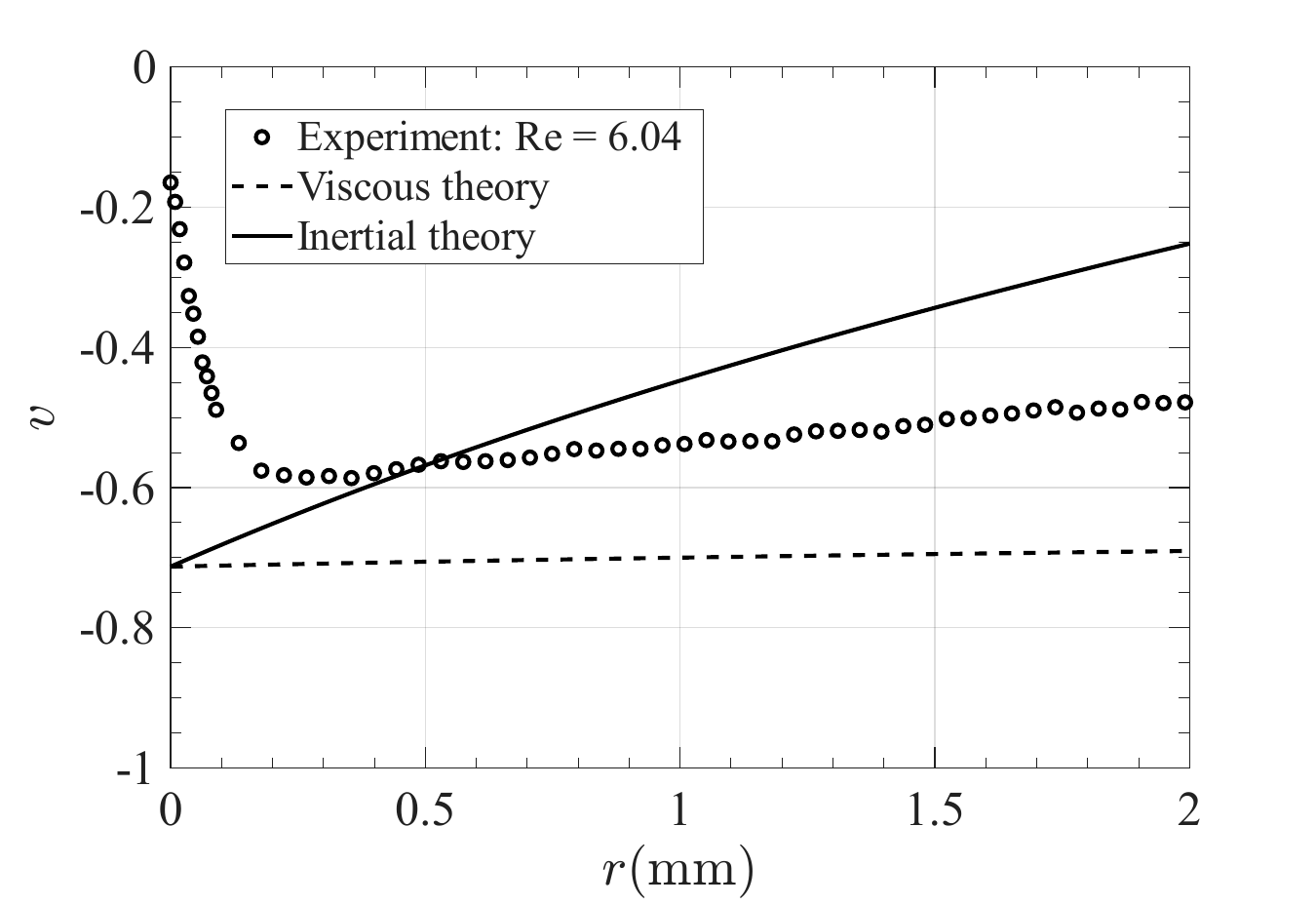}
    \caption{A variation of dimensionless interfacial speed with respect to the radial distance from the contact line for $Re$ = 6.04. The speed is normalized by the plate speed $U$. The ‘o’ markers denote experimental data, the dashed curve represents the prediction from viscous theory, and the black solid curve shows the prediction from inertial theory.}
\label{fig:interfacial_viscous_inertia}
\end{figure}
We can incorporate curvature effects into the interfacial speed by inculcating the angular wedge angle $\beta$ similar to the equn.~\ref{eq:interfacial_speed_MWS}. The speed is therefore given by the combination of the radial and azimuthal velocity components:
\begin{equation}\label{eq:interfacial_inertia_MWS}
    v_{i}^{i-MWS} = u_r(r,\beta)\cos(\alpha - \beta) 
+ u_\theta(r,\beta)\sin(\alpha - \beta),
\end{equation}
where $u_r|_{\theta=\beta}$ and $u_{\theta}|_{\theta=\beta}$ can be expressed as:
\begin{equation}
\begin{aligned}
u_r|_{\theta=\beta} =\;& f_1^{'} + \frac{r}{\nu}f_2^{'}; \hspace{4mm} \\ 
u_{\theta}|_{\theta=\beta} =& -f_1 - r\frac{\partial f_1}{\partial \beta}\frac{\partial \beta}{\partial r} -\frac{2r}{\nu}f_2 - \frac{r^2}{\nu}\frac{\partial f_2}{\partial \beta}\frac{\partial \beta}{\partial r}.
\end{aligned}
\end{equation}

Here, $\alpha$ and $\beta$, as shown in Figure~\ref{fig:curve_interface}, represent the local slope of the interface and the angular position measured from the contact line, respectively. The velocity components, $u_r|_{\theta=\beta}$ and $u_{\theta}|_{\theta=\beta}$, will be calculated numerically and will be substituted in equn.~\ref{eq:interfacial_inertia_MWS} to determine $v_i^{i-MWS}$. A representative result is shown in Figure~\ref{fig:interfacial_viscous_inertia}, illustrating that inertia causes the interfacial speed to decrease with radial distance and depend on the Reynolds number.
\section{Numerical framework}\label{apx:numerics}
\begin{figure}
    \centering
        \includegraphics[trim = 0mm 65mm 0mm 60mm, clip, width=0.5\textwidth]{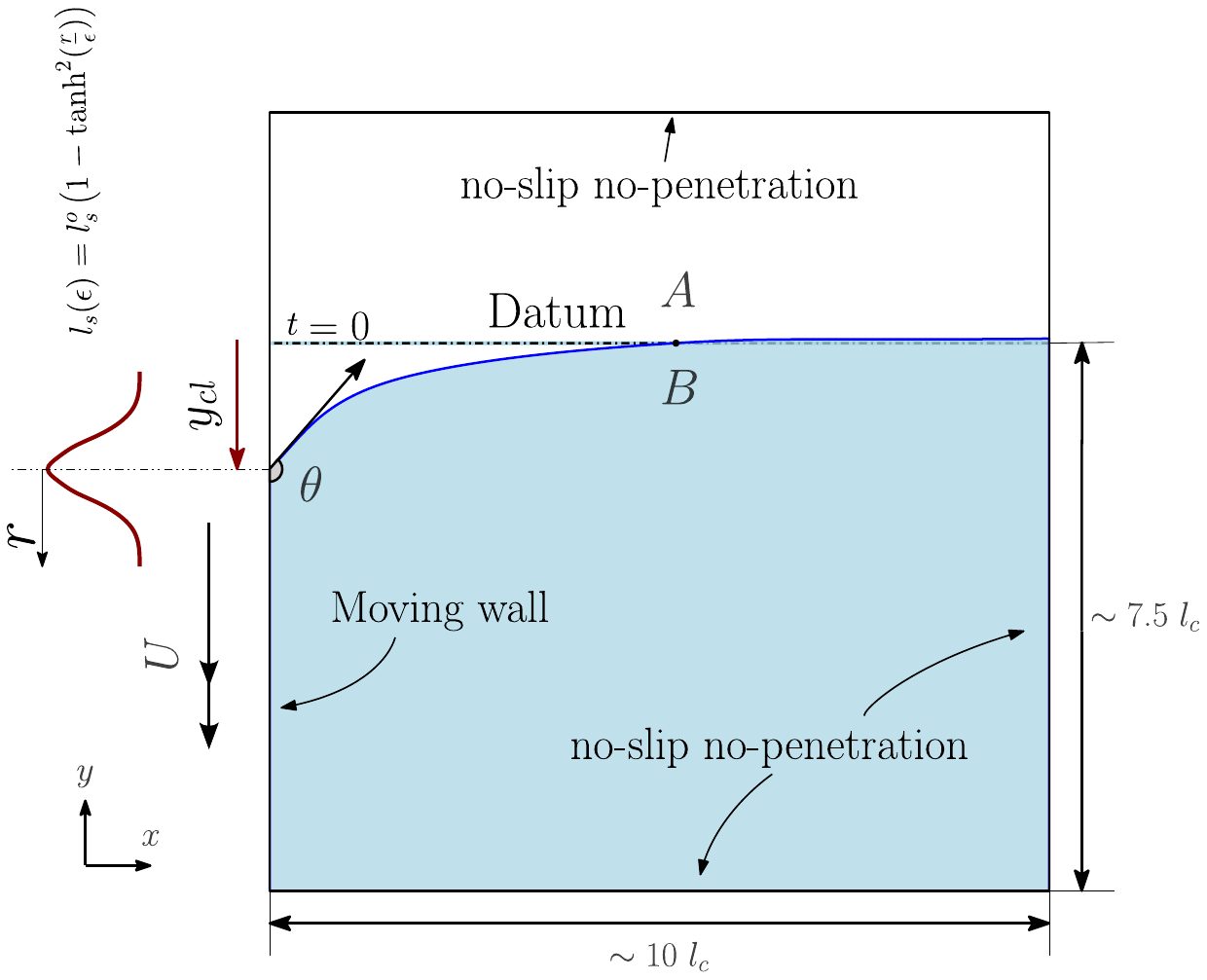}
    \caption{Schematic showing the numerical setup with boundary conditions. }
    \label{fig:Numerical-setup}
\end{figure}

We numerically set up a plate advancing experiment on open source software, \textit{Basilisk} \cite{popinet2015collaborators}. Figure~\ref{fig:Numerical-setup} shows a schematic describing the geometry and boundary conditions used in the simulations. For example, the left wall is vertically moving down with a spatially decaying slip centered at the contact line. The spatially decaying slip model reverts to the no-slip condition at distances greater than $\sim O(\epsilon)$. The slip model is explained in detail in our earlier study\cite{gupta2025investigation}. The remaining boundaries are fixed with no-slip and no-penetration boundary conditions. To eliminate wall effects due to higher plate speeds, the domain width is fixed to be $\sim 10~l_c$ wide, and the gas-fluid interface at $t=0$ is a flat line at a distance of $\sim 7.5~l_c$ from the bottom wall, which is allowed to evolve until the steady state is reached. 
The numerical cases are setup using the fluid properties and operating conditions similar to those used in experiments (see Tables~\ref{tab:properties_fluids} and \ref{tab:operating_parameters}). We prescribed the dynamic contact angle at the contact line similar to those observed in the experiments as reported in Table~\ref{tab:operating_parameters}. The contact angle boundary condition is implemented following the method of Afkhami \textit{et al.} \cite{afkhami2008height}. The grid size is chosen to match the experimental pixel resolution to enable a consistent comparison between experiments and numerical simulations.

%


\bibliography{mybib}

@article{huh1971hydrodynamic,
  title={Hydrodynamic model of steady movement of a solid/liquid/fluid contact line},
  author={Huh, Chun and Scriven, Laurence E},
  journal={J. Colloid Interface Sci.},
  volume={35},
  number={1},
  pages={85--101},
  year={1971},
  publisher={Elsevier}
}

@article{cox1986dynamics,
  title={The dynamics of the spreading of liquids on a solid surface. Part 1. Viscous flow},
  author={Cox, RG},
  journal={J. Fluid Mech.},
  volume={168},
  pages={169--194},
  year={1986},
  publisher={Cambridge University Press}
}

@article{chen1997velocity,
  title={The velocity field near moving contact lines},
  author={Chen, Q and Ram{\'e}, E and Garoff, S},
  journal={J. Fluid Mech.},
  volume={337},
  pages={49--66},
  year={1997},
  publisher={Cambridge University Press}
}

@article{dussan1979spreading,
  title={On the spreading of liquids on solid surfaces: static and dynamic contact lines},
  author={Dussan, EB},
  journal={Annual Review of Fluid Mechanics},
  volume={11},
  number={1},
  pages={371--400},
  year={1979},
  publisher={Annual Reviews 4139 El Camino Way, PO Box 10139, Palo Alto, CA 94303-0139, USA}
}

@article{blake1969kinetics,
  title={Kinetics of liquidliquid displacement},
  author={Blake, TD and Haynes, JM},
  journal={J. Colloid Interface Sci.},
  volume={30},
  number={3},
  pages={421--423},
  year={1969},
  publisher={Elsevier}
}

@article{shikhmurzaev1993moving,
  title={The moving contact line on a smooth solid surface},
  author={Shikhmurzaev, Yu D},
  journal={Int. J. Multiph. Flow},
  volume={19},
  number={4},
  pages={589--610},
  year={1993},
  publisher={Elsevier}
}

@article{shikhmurzaev1997moving,
  title={Moving contact lines in liquid/liquid/solid systems},
  author={Shikhmurzaev, Yulii D},
  journal={J. Fluid Mech.},
  volume={334},
  pages={211--249},
  year={1997},
  publisher={Cambridge University Press}
}

@article{voinov1976hydrodynamics,
  title={Hydrodynamics of wetting},
  author={Voinov, OV},
  journal={Fluid Dyn.},
  volume={11},
  number={5},
  pages={714--721},
  year={1976},
  publisher={Springer}
}

@article{kirkinis2014moffatt,
  title={Moffatt vortices induced by the motion of a contact line},
  author={Kirkinis, E and Davis, SH},
  journal={J. Fluid Mech.},
  volume={746},
  year={2014},
  publisher={Cambridge University Press}
}

@article{kirkinis2013hydrodynamic,
  title={Hydrodynamic theory of liquid slippage on a solid substrate near a moving contact line},
  author={Kirkinis, E and Davis, Stephen H},
  journal={Physical Review Letters},
  volume={110},
  number={23},
  pages={234503},
  year={2013},
  publisher={APS}
}

@article{febres2017existence,
  title={Existence of Moffatt vortices at a moving contact line between two fluids},
  author={Febres, Mijail and Legendre, Dominique},
  journal={Phys. Rev. Fluid},
  volume={2},
  number={11},
  pages={114002},
  year={2017},
  publisher={APS}
}

@article{snoeijer2006free,
  title={Free-surface flows with large slopes: Beyond lubrication theory},
  author={Snoeijer, Jacco H},
  journal={Physics of Fluids},
  volume={18},
  number={2},
  year={2006},
  publisher={AIP Publishing}
}

@article{chan2013hydrodynamics,
  title={Hydrodynamics of air entrainment by moving contact lines},
  author={Chan, Tak Shing and Srivastava, S and Marchand, A and Andreotti, B and Biferale, L and Toschi, F and Snoeijer, Jacobus Hendrikus},
  journal={Physics of fluids},
  volume={25},
  number={7},
  year={2013},
  publisher={AIP Publishing}
}

@article{chan2020cox,
  title={Cox--Voinov theory with slip},
  author={Chan, Tak Shing and Kamal, Catherine and Snoeijer, Jacco H and Sprittles, James E and Eggers, Jens},
  journal={Journal of fluid mechanics},
  volume={900},
  pages={A8},
  year={2020},
  publisher={Cambridge University Press}
}

@article{snoeijer2013moving,
  title={Moving contact lines: scales, regimes, and dynamical transitions},
  author={Snoeijer, Jacco H and Andreotti, Bruno},
  journal={Annual review of fluid mechanics},
  volume={45},
  pages={269--292},
  year={2013},
  publisher={Annual Reviews}
}

@article{gupta2023,
  title={An experimental study of flow near an advancing contact line: a rigorous test of theoretical models},
  author={Gupta, Charul and Choudhury, Anjishnu and Chandrala, Lakshmana D and Dixit, Harish N},
  journal={Journal of Fluid Mechanics},
  volume={1000},
  pages={A45},
  year={2024},
  publisher={Cambridge University Press}
}

@article{gupta2024experimental,
  title={An experimental investigation of flow fields near a liquid--liquid moving contact line},
  author={Gupta, Charul and Chandrala, Lakshmana D and Dixit, Harish N},
  journal={The European Physical Journal Special Topics},
  volume={233},
  number={8},
  pages={1653--1663},
  year={2024},
  publisher={Springer}
}

@article{hoffman1975study,
  title={A study of the advancing interface. I. Interface shape in liquid—gas systems},
  author={Hoffman, Richard L},
  journal={Journal of colloid and interface science},
  volume={50},
  number={2},
  pages={228--241},
  year={1975},
  publisher={Elsevier}
}

@article{le2005shape,
  title={Shape and motion of drops sliding down an inclined plane},
  author={Le Grand, Nolwenn and Daerr, Adrian and Limat, Laurent},
  journal={J. Fluid Mech.},
  volume={541},
  pages={293--315},
  year={2005},
  publisher={Cambridge University Press}
}

@article{rio2005boundary,
  title={Boundary conditions in the vicinity of a dynamic contact line: experimental investigation of viscous drops sliding down an inclined plane},
  author={Rio, E and Daerr, A and Andreotti, B and Limat, L},
  journal={Physical review letters},
  volume={94},
  number={2},
  pages={024503},
  year={2005},
  publisher={APS}
}

@article{puthenveettil2013motion,
  title={Motion of drops on inclined surfaces in the inertial regime},
  author={Puthenveettil, Baburaj A and Senthilkumar, Vijaya K and Hopfinger, EJ},
  journal={J. Fluid Mech.},
  volume={726},
  pages={26--61},
  year={2013},
  publisher={Cambridge University Press}
}

@article{duez2007making,
  title={Making a splash with water repellency},
  author={Duez, Cyril and Ybert, Christophe and Clanet, Christophe and Bocquet, Lyderic},
  journal={Nature physics},
  volume={3},
  number={3},
  pages={180--183},
  year={2007},
  publisher={Nature Publishing Group UK London}
}

@book{shikhmurzaev2007capillary,
  title={Capillary flows with forming interfaces},
  author={Shikhmurzaev, Yulii D},
  year={2007},
  publisher={Chapman and Hall/CRC}
}

@article{varma2021inertial,
  title={Inertial effects on the flow near a moving contact line},
  author={Varma, Akhil and Roy, Anubhab and Puthenveettil, Baburaj A},
  journal={J. Fluid Mech.},
  volume={924},
  pages={A36},
  year={2021},
  publisher={Cambridge University Press}
}

@article{hancock1981effects,
  title={Effects of inertia in forced corner flows},
  author={Hancock, C and Lewis, E and Moffatt, HK},
  journal={J. Fluid Mech.},
  volume={112},
  pages={315--327},
  year={1981},
  publisher={Cambridge University Press}
}

@phdthesis{he_eindhoven2020,
    author = {He, B.},
    title = {Formation of watermark defects during immersion lithography},
    school = {Technische Universiteit Eindhoven},
    year = {2020} 
}

@article{dussan1991,
title={On identifying the appropriate boundary conditions at a moving contact line: an experimental investigation},  
author={Dussan V., E. B. and Ramé, Enrique and Garoff, Stephen},
journal={J. Fluid Mech.},
volume={230},
pages={97–116},
year={1991},
publisher={Cambridge University Press},
}

@article{stoev1999effects,
  title={Effects of inertia on the hydrodynamics near moving contact lines},
  author={Stoev, K and Ram{\'e}, E and Garoff, S},
  journal={Physics of Fluids},
  volume={11},
  number={11},
  pages={3209--3216},
  year={1999},
  publisher={American Institute of Physics}
}

@misc{popinet2015collaborators,
  title={collaborators. Basilisk},
  author={Popinet, St{\'e}phane},
  year={2015}
}

@article{afkhami2008height,
  title={Height functions for applying contact angles to 2D VOF simulations},
  author={Afkhami, Shahriar and Bussmann, Markus},
  journal={International journal for numerical methods in fluids},
  volume={57},
  number={4},
  pages={453--472},
  year={2008},
  publisher={Wiley Online Library}
}

@article{fullana2024consistent,
  title={A consistent treatment of dynamic contact angles in the sharp-interface framework with the generalized Navier boundary condition},
  author={Fullana, Tomas and Kulkarni, Yash and Fricke, Mathis and Popinet, St{\'e}phane and Afkhami, Shahriar and Bothe, Dieter and Zaleski, St{\'e}phane},
  journal={arXiv preprint arXiv:2411.10762},
  year={2024}
}

@article{cox1998inertial,
  title={Inertial and viscous effects on dynamic contact angles},
  author={Cox, RG},
  journal={Journal of Fluid Mechanics},
  volume={357},
  pages={249--278},
  year={1998},
  publisher={Cambridge University Press}
}

@article{vandre2014characteristics,
  title={Characteristics of air entrainment during dynamic wetting failure along a planar substrate},
  author={Vandre, E and Carvalho, MS and Kumar, S},
  journal={Journal of fluid mechanics},
  volume={747},
  pages={119--140},
  year={2014},
  publisher={Cambridge University Press}
}

@article{gupta2025investigation,
  title={An Investigation of Flow and Interface Dynamics Near a Moving Contact Line at Obtuse Contact Angles},
  author={Gupta, Charul and Sangadi, Venkata Sai Anvesh and Chandrala, Lakshmana Dora and Dixit, Harish N},
  journal={arXiv preprint arXiv:2502.09953},
  year={2025}
}

\end{document}